\documentstyle[12pt]{article}

\def\hybrid{\topmargin 0pt      \oddsidemargin 0pt
        \headheight 0pt \headsep 0pt
        \voffset=-0.5cm
        \textwidth 6.5in        
        \textheight 9in         
        \marginparwidth 0.0in
        \parskip 5pt plus 1pt   \jot = 1.5ex}
\catcode`\@=11
\def\marginnote#1{}

\newcount\hour
\newcount\minute
\newtoks\amorpm
\hour=\time\divide\hour by60
\minute=\time{\multiply\hour by60 \global\advance\minute by-\hour}
\edef\standardtime{{\ifnum\hour<12 \global\amorpm={am}%
        \else\global\amorpm={pm}\advance\hour by-12 \fi
        \ifnum\hour=0 \hour=12 \fi
        \number\hour:\ifnum\minute<10 0\fi\number\minute\the\amorpm}}
\edef\militarytime{\number\hour:\ifnum\minute<10 0\fi\number\minute}

\def\draftlabel#1{{\@bsphack\if@filesw {\let\thepage\relax
   \xdef\@gtempa{\write\@auxout{\string
      \newlabel{#1}{{\@currentlabel}{\thepage}}}}}\@gtempa
   \if@nobreak \ifvmode\nobreak\fi\fi\fi\@esphack}
        \gdef\@eqnlabel{#1}}
\def\@eqnlabel{}
\def\@vacuum{}
\def\draftmarginnote#1{\marginpar{\raggedright\scriptsize\tt#1}}

\def\draftlabel#1{{\@bsphack\if@filesw {\let\thepage\relax
   \xdef\@gtempa{\write\@auxout{\string
      \newlabel{#1}{{\@currentlabel}{\thepage}}}}}\@gtempa
   \if@nobreak \ifvmode\nobreak\fi\fi\fi\@esphack}
        \gdef\@eqnlabel{#1}}
\def\@eqnlabel{}
\def\@vacuum{}
\def\draftmarginnote#1{\marginpar{\raggedright\scriptsize\tt#1}}

\def\draft{\oddsidemargin -.5truein
        \def\@oddfoot{\sl preliminary draft \hfil
        \rm\thepage\hfil\sl\today\quad\militarytime}
        \let\@evenfoot\@oddfoot \overfullrule 3pt
        \let\label=\draftlabel
        \let\marginnote=\draftmarginnote
   \def\@eqnnum{(\theequation)\rlap{\kern\marginparsep\tt\@eqnlabel}%
\global\let\@eqnlabel\@vacuum}  }

\def\numberbysection{\@addtoreset{equation}{section}
        \def\theequation{\thesection.\arabic{equation}}}

\def\underline#1{\relax\ifmmode\@@underline#1\else
        $\@@underline{\hbox{#1}}$\relax\fi}

\def\titlepage{\@restonecolfalse\if@twocolumn\@restonecoltrue\onecolumn
     \else \newpage \fi \thispagestyle{empty}\c@page\z@
        \def\thefootnote{\fnsymbol{footnote}} }

\def\endtitlepage{\if@restonecol\twocolumn \else  \fi
        \def\thefootnote{\arabic{footnote}}
        \setcounter{footnote}{0}}  
\relax

\numberbysection
\hybrid

\def\beq{\begin{equation}}
\def\eeq{\end{equation}}
\def\p{\partial}
\def\G{\Gamma}
\def\g{\gamma}
\def\s{\sigma}

\def\a{\alpha}
\def\b{\beta}

\def\M{{\cal M}}
\def\N{{\cal N}}

\def\res{{\rm res}}

\def \l{\lambda}

\def \matrix #1 {\left(\begin{array}{cc} #1 \end{array}\right)}

\newtheorem{lem}{Lemma}[section]

\begin{document}

\begin{titlepage}
\title{\bf Spin Chain Models with Spectral Curves from M Theory}

\bigskip

\author{I.Krichever\thanks{Supported
in part by the National Science Foundation under
grant DMS-98-02577}
\ \, and D.H.Phong\thanks{Supported in
part by the National Science Foundation under
grant DMS-98-00783}}
\date{December 20, 1999}

\maketitle

\centerline{${}^*$ Department of Mathematics}
\centerline{Columbia University}
\centerline{New York, NY 10027}
\centerline{and}
\centerline{Landau Institute of Theoretical Physics}
\centerline{Kosygina str. 2, 117940 Moscow, Russia}
\centerline{email: krichev@math.columbia.edu}

\bigskip

\centerline{${}^\dagger$ Department of Mathematics}
\centerline{Columbia University}
\centerline{New York, NY 10027}
\centerline{email: phong@math.columbia.edu}

\bigskip


\bigskip

\centerline{\bf Abstract}

\bigskip

We construct the integrable model corresponding to
the $\N=2$ supersymmetric $SU(N)$ gauge theory with matter
in the antisymmetric representation, using the spectral curve
found by Landsteiner and Lopez through M Theory. The model turns out to
be the Hamiltonian reduction of a $N+2$ periodic spin chain model,
which is Hamiltonian with respect to
the universal symplectic form we had constructed earlier
for general soliton equations in the Lax or Zakharov-Shabat representation.


\vfill

\end{titlepage}
\newpage

\section{Introduction}

The main goal of this paper is to construct the integrable model
which corresponds to the $\N=2$ SUSY $SU(N)$ Yang-Mills theory
with a hypermultiplet in the antisymmetric representation.
The 1994 work of Seiberg and Witten \cite{seiberg}
had shown that the Wilson effective
action of $\N=2$ SUSY Yang-Mills theory is determined by
a fibration of spectral curves $\Gamma$
equipped with a meromorphic one-form $d\lambda$, now known as
the Seiberg-Witten differential. It was soon recognized afterwards
\cite{survey1, martinec, donagi} that this set-up is indicative of
an underlying integrable model, with the vacuum moduli of the
Yang-Mills theory corresponding to the action variables
of the integrable model. In fact, in the special case of hyperelliptic
curves, a similar set-up for the construction of action
variables as periods of a meromorphic differential
had been introduced in
\cite{novikov}. This unexpected relation between $\N=2$
Yang-Mills theories on one hand and integrable models
has proven to be very beneficial for both sides.
The Seiberg-Witten differential has led to
a universal symplectic form for soliton equations
in the Lax or Zakharov-Shabat representation
\cite{kp1, kp2}. The connection with integrable models has
helped solve the $SU(N)$ Yang-Mills theory with a hypermultiplet
in the adjoint representation \cite{donagi, dhoker1}, as well as pure
Yang-Mills theories with arbitrary simple gauge groups $G$
\cite{martinec}. Conversely,
the connection with Yang-Mills theories has led to new
integrable models, such as the twisted Calogero-Moser systems
associated with Yang-Mills theories with non-simply laced
gauge group and matter in the adjoint representation \cite{dhoker2},
and the elliptic analog of the Toda lattice \cite{el}.
\footnote{We refer to \cite{survey2, survey3} for more complete lists of
references.}

Despite all these successes, we still do not know at this moment
how to identify or construct the correct integrable model
corresponding to a given Yang-Mills theory. This is a serious drawback,
since the integrable model can be instrumental in investigating
key physical issues such as duality, the renormalization group, or instanton
corrections \cite{dkp, k1, wdvv}. At the same time, the list of spectral curves
continues to grow, thanks in particular to methods from M theory
\cite{witten, landsteiner} and geometric engineering \cite{katz}. It seems
urgent to develop methods which can identify the correct integrable
model from a given spectral curve and Seiberg-Witten differential.

In the case of interest in this paper, namely the $SU(N)$ gauge
theory with antisymmetric matter, the Seiberg-Witten differential
and spectral curve had been found by
Landsteiner and Lopez \cite{landsteiner} using branes and M theory. The
Seiberg-Witten differential $d\lambda$ is given by
\beq
d\lambda=x{dy\over y}. \label{sw1}
\eeq
The spectral curve is of the form
\beq
y^3-(3\Lambda^{N+2}+x^2\sum_{i=0}^Nu_ix^i)y^2+
(3\Lambda^{N+2}+x^2\sum_{i=0}^N(-)^iu_ix^i)\Lambda^{N+2}y-
\Lambda^{3(N+2)}=0,
\label{ll0}
\eeq
where $\Lambda$ is a renormalization scale.
For the $SU(N)$ gauge theories, one restricts to $u_{N}=1$, $u_{N-1}=0$,
so that the moduli dimension is $N-1$,
which is the rank of the gauge group $SU(N)$.
The Landsteiner-Lopez curve (\ref{ll0}) and differential (\ref{sw1})
have been studied extensively by Ennes, Naculich,
Rhedin, and Schnitzer \cite{schnitzer1}.
In particular, they have verified that the curve and differential
do reproduce the correct perturbative behavior of the prepotential
predicted by asymptotic freedom.
The problem which we wish to address here is the one of finding
a dynamical system which is integrable in the sense that it admits
a Lax pair, and which corresponds to the Landsteiner-Lopez curve
and Seiberg-Witten differential
(\ref{sw1}) in the sense that its spectral curve
is of the form (\ref{ll0}), and its action variables are the
periods of $d\lambda$ along $N-1$ suitable cycles on
$\Gamma$.

We have succeeded in constructing two integrable spin chain models,
whose spectral curves are given exactly by the Landsteiner-Lopez
curves. However, the action variables of the desired integrable model
must be given by $d\lambda=x{dy\over y}$, and here the two
models differ significantly. For one model, referred to as {\it the
odd divisor spin model}, the 2-form resulting from $d\lambda$ vanishes
identically. For the other, referred to as {\it the even divisor spin model},
the Hamiltonian reduction of the 2-form resulting from $d\lambda$
to the moduli space of vacua $\{u_N=1, \quad u_{N-1}=0\}$
is non-degenerate, and the reduced system is indeed Hamiltonian
with respect to this symplectic form,
with Hamiltonian $H=u_{N-2}$. Thus the latter model is the integrable
system we are looking for.

Our main result is as follows\footnote{The notation is
explained in greater detail in \S 3 and \S 5.}. Let $q_n, p_n$ be 3-dimensional vectors
which are $N+2$ periodic, i.e. $p_{n+N+2}=p_n$, $q_{n+N+2}=q_n$, and
satisfy the constraints
\begin{eqnarray}\label{con1}
p_n^Tq_n&=&0 \\
p_n=g_0p_{-n-1}&, &
q_n=g_0q_{-n-1} \label{con}
\end{eqnarray}
where $g_0$ is the diagonal matrix
\beq\label{G}
g_0=\left(\begin{array}{ccc}
1 & 0 & 0 \\
0 & -1 & 0 \\
0 & 0 & 1
\end{array}\right).
\eeq
Consider the dynamical system
\beq
\dot p_n={p_{n+1}\over p_{n+1}^Tq_{n}}+{p_{n-1}\over p_{n-
1}^Tq_{n}}+\mu_np_n,\ \
\dot q_n=-{q_{n+1}\over p_{n}^Tq_{n+1}}-{q_{n-1}\over p_{n}^Tq_{n-1}}-
\mu_np_n.
\label{pq0}
\eeq
for some scalar functions $\mu_n(t)$. The system is invariant under
the gauge group $G$ generated by the following gauge
transformations
\begin{eqnarray}\label{g1}
p_n\to \lambda_n p_n &,& q_n\to \lambda_n^{-1}q_n, \\ \label{g2}
p_n\to W^Tp_n &,& q_n\to W^{-1}q_n,
\end{eqnarray}
Here $W$ is a $3\times 3$ matrix which commutes with $g_0$,
$Wg_0=g_0W$.
Define the $3\times 3$ matrices $L(x)$ and $M(x)$ by
\beq
L(x)=\prod_{n=0}^{N+1}(1+x q_np_n^T),\ \ \
M(x)=x\left({q_{N+1}p_0^T\over p_0^Tq_{N+1}}-{q_0p_{N+1}^T\over
p_{N+1}^Tq_0}\right)
\eeq

\bigskip
\noindent
{\bf Main Theorem.} $\bullet$ {\it The dynamical system {\rm(\ref{pq0})} is
equivalent
to the Lax equation}
\beq
\dot L(x)=[M(x),L(x)];
\label{lax}
\eeq
$\bullet$ {\it The spectral curves $\Gamma=\{R(x,y)\equiv{\rm det}\big(yI-
L(x)\big)=0\}$ are invariant under the flow {\rm (\ref{pq0})}, and are exactly
the curves of the Landsteiner-Lopez form (\ref{ll0}) (with $\Lambda^{N+2}$
normalized to $1$)};\hfil\break
$\bullet$ {\it There is a natural map $(q_n,p_n)\to (\Gamma,D)$
from the space of all spin chains satisfying the constraints
{\rm (\ref{con1},\ref {con})} to the space of pairs $(\Gamma, D)$,
where $\Gamma$ is a Landsteiner-Lopez curve, and $D=\{z_1,\cdots,
z_{2N+1}\}$ is a divisor whose class $[D]=[D^{\sigma}]$ is symmetric under
the involution}
\beq
\sigma: \quad (x,y)=z\to z^\sigma=(-x,y^{-1}).\label{sigma}
\eeq
{\it For a given $(q_n,p_n)$,
$D$ is the set of poles of the Bloch function $\psi_0$,
$L(x)\psi_0=y\psi_0(x)$;}
\hfil\break
$\bullet$ {\it Let ${\cal M}_0$ be the space of pairs $\{\Gamma, [D]\}$,
where $\Gamma$ is a Landsteiner-Lopez curve
with $u_N=1$, $u_{N-1}=0$, and $[D]$ is a
divisor class which is symmetric under the involution $\sigma$.
Then the space $\M_0$ has dimension $2(N-1)$.
The map $(q_n,p_n)\to (\Gamma,D)$ descends to a map
between the two spaces}
\beq
\{(q_n,p_n)\}/G \leftrightarrow {\cal M}_0,\label{iso}
\eeq
{\it where on the left hand side, we have factored out
the gauge group $G$ from the space of periodic spin chains
satisfying the constraints {\rm (\ref{con1},\ref{con})}.
At a generic curve $\Gamma$ and a divisor $[D]$ in general position,
the map {\rm (\ref{iso})} is a local isomorphism.}
\hfil\break
$\bullet$ {\it Let the action variables $a_i$ and the angle variables $\phi_i$
be defined on the space $\M_0$ by}
\beq
a_i=\oint_{A_i}d\lambda,
\quad\quad
\phi_i=\sum_{i=1}^{2N+1}\int^{z_i}d\omega_i
\eeq
{\it where $\{A_i\}_{1\leq i\leq N-1}$ and $\{d\omega_i\}_{1\leq i\leq N-1}$, are
respectively a basis for the even cycles and a basis
for the even holomorphic differentials on $\Gamma$. Then}
\beq
\omega=\sum_{i=1}^{N-1}\delta a_i\wedge\delta\phi_i\label{sf0}
\eeq
{\it defines a symplectic form on the $2(N-1)$-dimensional space $\M_0$};
\hfil\break
$\bullet$
{\it The dynamical system {\rm (\ref{pq0})} is Hamiltonian with respect the
symplectic form
{\rm (\ref{sf0})}. The Hamiltonian is $H=u_{N-2}$.}

\bigskip

In terms of the $(q_n,p_n)$ dynamical variables,
the Hamiltonian can be expressed under the form
\beq\label{H12}
H={u_{N-2}\over u_N}-{u^2_{N-1}\over 2u^2_N}=\sum_{n=0}^{N+1}
{(p_n^Tq_{n-3})\over (p_n^Tq_{n-1})(p_{n-1}^Tq_{n-2})(p_{n-2}^Tq_{n-3})}-
{(p_n^Tq_{n-2})^2\over 2(p_n^Tq_{n-1})^2(p_{n-1}^Tq_{n-2})^2},
\eeq
where we have used the constraint $u_N=1$, $u_{N-1}=0$
to write $H$ as $H={u_{N-2}\over u_N}-{u_{N-1}^2\over 2 u_N^2}$.

\bigskip

We would like to note the similarity of the Lax matrix $L$ in (1.9)
to the $2\times 2$ Lax matrix used in \cite{korchem}
for the integration of a quasi-classical approximation to a 
system of reggeons in $QCD$.

\bigskip

A key tool in our analysis is the construction
of \cite{kp1, kp2}, which shows that symplectic forms
constructed in terms of Seiberg-Witten differentials
can also be constructed directly
in terms of the Lax representation of integrable models.
The latter are given by the following universal formula
\cite{kp1, kp2}
\beq
\omega={1\over 2}
\sum_{\a}{\rm Res}_{P_{\alpha}}<\psi_{n+1}^*\delta L_n(x)\wedge
\delta\psi_n>dx
\eeq
where $\psi_n$ and $\psi_{n+1}^*$ are the Bloch and dual Bloch functions
of the system, and $P_{\a}$ are marked punctures on the spectral curve $\Gamma$.
In the present case, $P_{\a}$ are the 3 points on $\Gamma$ above $x=\infty$.

\medskip

Finally, we note that the odd divisor spin model (which we describe in \S 3.1
and \S 6) may be of independent interest. Although the symplectic
form associated to the Seiberg-differential $x{dy\over y}$ is
degenerate in this case, the model does admits a Hamiltonian structure with
non-degenerate symplectic form, but one which is associated rather with
the form $d\lambda_{(1)}=\ln y {dx\over x}$. As suggested in \cite{5d}, the form
$\ln y{dx\over x}$ is also indicative of supersymmetric Yang-Mills
theories, but in 5 or 6 dimensions with $\N=1$ supersymmetry.

\section{Geometry of the Landsteiner-Lopez curve}

We begin by identifying the geometric features of the generic
Landsteiner-Lopez curve which will play an important role in the
sequel. Fixing the normalization $\Lambda^{N+2}=1$, we can write
\beq
\Gamma: \quad R(x,y)\equiv y^3-f(x)y^2+f(-x)y-1=0\label{ll1}
\eeq
where $f(x)$ is a polynomial of the form
\beq
f(x)=3+x^2P_N(x),\quad P_N(x)=\sum_{i=0}^Nu_ix^i\label{ll2}
\eeq
The parameters $u_0,\cdots,u_N$ are
the moduli of the Landsteiner-Lopez curve.

\medskip

$\bullet$ The Landsteiner-Lopez curve $\Gamma$ is a three-fold covering
of the complex plane in the $x$ variable. It is invariant under
the involution $\sigma$ defined in (\ref{sigma}). The important points
on $\Gamma$ are the singular points, the points above $x=\infty$,
and the branch points. We discuss now all these points in turn.

\medskip

$\bullet$ The singular points are the points where
\beq
\partial_xR(x,y)=\partial_yR(x,y)=0
\eeq
The generic Landsteiner-Lopez curve has exactly one
singular point, namely $(x,y)=(0,1)$.
At this point, the equation (\ref{ll1}) has a triple
root, and all three sheets of the curve intersect. For generic
values of the moduli $u_i$,
all three solutions $y$ of $R(x,y)=0$
can be expressed as power series in $x$ in a neighborhood of $x=0$
\beq
y(x)=1+\sum_{i=1}^{\infty} y_ix^i \label{100}
\eeq
In fact, we can substitute (\ref{100}) into (\ref{ll1})
to find recursively all coefficients $y_i$, with the first
coefficient $y_1$ a solution of
\beq\label{101}
y_1^3-u_0y_1+2u_1=0.
\eeq
For generic $u_0$, $u_1$, this equation does admit three distinct
solutions for $y_1$, which lead in turn to the three distinct
solutions. These three distinct solutions
provide effectively a smooth resolution of the curve $\Gamma$,
where the crossing point $y=1$ above $x=0$ has been separated
into 3 distinct points $Q_{\alpha}$, $1\leq\alpha\leq 3$.
Under the involution $\sigma$, the leading terms in the
three solutions (\ref{100}) transform as
\beq
(x,1+y_1x+\cdots)
\to (-x, (1-y_1x+\cdots)^{-1})
=(-x,1+y_1x+\cdots)
\eeq
Since the three solutions $y_1$ of the equation (\ref{101})
are distinct for generic values of the moduli $u_i$, we see that
each of the three points $Q_{\alpha}$ above $x=0$ are
fixed under the involution $\sigma$.

\medskip

$\bullet$ For generic values of the moduli $u_i$,
there are also three distinct branches of $y(x)$ near $x=\infty$.
A first branch $y(x)=O(x^{N+2})$ with a pole of order
$N+2$ can be readily found
\beq\label{y27}
y(x)=x^{N+2}(u_N+u_{N-1}x^{-1}+u_{N-2}x^{-2}+\cdots).
\eeq
(The first three coefficients in $y(x)$ turn out
to be exactly the first three coefficients $u_N,u_{N-1}$ and
$u_{N-2}$ in the polynomial $P_N(x)$ of (\ref{ll2}).)
We denote by $P_1$ the corresponding point above $x=\infty$.
In view of the involution $\sigma$, a second branch $y(x)=
O(x^{-(N+2)})$ with a zero of order $N+2$ exists which
is the image of the first branch under $\sigma$
\beq\label{y28}
y(x)=(-x)^{-(N+2)}{1\over u_N}
\left(1+{u_{N-1}\over u_N}x^{-1}+{u_{N-1}^2-u_Nu_{N-2}\over u_N^2}x^{-2}
+\cdots\right)
\eeq
The corresponding point above $x=\infty$ is denoted $P_3$.
Finally, the involution $\sigma$ implies that the third branch
$y(x)$ is regular and fixed under $\sigma$
\beq
y(x)=(-)^{N+2}\big[1+O\left({1\over x}\right)\big]
\eeq
Denoting the corresponding point above $x=\infty$ by $P_2$, we have
\beq
\sigma: P_1\leftrightarrow P_3,\quad \sigma: P_2\leftrightarrow P_2.
\eeq

\medskip

$\bullet$
The branching points of $\G$ over $x$-plane
are just the zeroes on $\G$ of the function $\partial_yR(x,y)$
which are different from the singular points
$Q_{\a}$. This function has a pole of order $2(N+2)$ at $P_1$ and
a pole of order $(N+2)$ at each of the points
$P_2$ and $P_3$. Therefore, it has $4N+8$ zeros. At each of the
points $Q_{\a}$ the function $\partial_yR(x,y)$ has zeros of order 2.
Hence
\beq
\#\{ {\rm Branch\ Points}\}=4N+2.
\eeq
Note that for generic moduli $u_i$,
neither $0$ nor $\infty$ is a branch point, in view of
our previous discussion. Also for generic $u_i$, we can assume
that the ramification index at all branch points is $2$.
Thus the total branching number is just the number of branch points.
Since the number of sheets is $3$, the Riemann-Hurwitz formula
can be written as $g(\Gamma)=-3+{1\over 2}(4N+2)+1$ in this case.
Thus the genus $g(\Gamma)$ of the curve $\Gamma$ is
\beq
g(\Gamma)=2N-1.\label{genus}
\eeq

\medskip

$\bullet$ For generic moduli $u_i$, the involution $\s:\G\to \G$ has
exactly four fixed points, namely the three points $Q_{\alpha}$
above $x=0$ and the point $P_2$ above $x=\infty$. That implies that
the factor-curve $\G/\s$ has genus
\beq
g(\Gamma/\sigma)=N-1.
\eeq
The involution $\s$ induces an involution of the Jacobian variety $J(\G)$ of
$\G$. The odd part $J^{Pr}(\G)$ of $J(\G)$ is the Prym variety
and the even part is isogenic to the Jacobian $J(\G/\sigma)$ of the
factor-curve $\G/\sigma$.
The dimension of the space of divisors $[D]$
which are even under $\sigma$ is equal to ${\rm dim}\,J(\G/\sigma)
=N-1$.

\section{The Spin Models}

We introduce two systems with the same family of
spectral curves (\ref{ll1}). One system has non-trivial dynamics along
the even while the other system has non-trivial dynamics
along the odd (Prym) directions of the Jacobian.
The system corresponding to the $SU(N)$ Yang-Mills theory
with a hypermultiplet in the anti-symmetric representation
is the even system.
We sketch here the outline of the construction of both models, leaving
the full discussion to sections \S 4-5.

Both models are periodic spin chain models, with a 3-dimensional
complex vector at each site. We view three-dimensional vectors $s$
as column vectors, with components $s_{\alpha}$, $1\leq\alpha\leq 3$.
We denote by $s^T$ the transpose of $s$, which is then a three-dimensional
row vector, with components $s^{\alpha}$. In particular, $s^Ts$ is a scalar,
while $ss^T$ is a $3\times 3$ matrix.
Since the odd divisor spin model is simpler,
we begin with it.

\subsection{The Odd Divisor Spin Model}

The odd divisor spin model is a $(N+2)$-periodic chain of complex
three-dimensional vectors
$s_n=s_{N+n+2},\
s_n=(s_{n,\a}),\ \a=1,2,3,$ subject to the constraint
\beq\label{con0}
s_n^Ts_n=\sum_{\a=1}^3s_n^{\a}s_{n,\a}=0,
\eeq
and the following equations of motion
\beq\label{s}
\dot s_n={s_{n+1}\over s_{n+1}^Ts_n}-{s_{n-1}\over s_{n-1}^Ts_n}.
\eeq
The constraint (\ref{con0}) and the equations of motion are invariant
under transformation of the spin chain by a matrix $V$ satisfying
the condition $V^TV=I$
\beq\label{gaugeodd}
s_n \to Vs_n
\eeq
The odd divisor spin model is integrable in the sense that the equations
of motion are equivalent to a Lax pair. To see this, we define the $3\times 3$
matrices $L_n(x)$ and $M_n(x)$ by
\begin{eqnarray}
L_n(x)& = & 1+x\,s_ns_n^T\label{laxoddl}\\
M_n(x)& = & x{1\over s_n^Ts_{n-1}}(s_{n-1}s_n^T+s_ns_{n-1}^T)\label{laxodd}
\end{eqnarray}
Then the compatibility condition for the system of equations
\begin{eqnarray}
\psi_{n+1}&=&L_n(x)\psi_n\\
\dot\psi_n &= & M_n(x)\psi_n
\end{eqnarray}
is given by
\beq
\dot L_n(x)=M_{n+1}(x)L_n(x)-L_n(x)M_n(x)\label{laxn}
\eeq
A direct calculation shows that for $L_n(x)$ and $M_n(x)$
defined as in (\ref{laxodd}), this equation is equivalent to
the equations of motion (\ref{s}) for the spin model.
Define now the monodromy matrix $L(x)$ by
\beq
L(x)=L_{N+1}(x)\cdots L_0(x)=\prod_{n=0}^{N+1}L_n(x)
\eeq
where the ordering in the product on the right hand side starts
by convention with the lowest indices on the right. Then $L(x)$ and
$M(x)=M_0(x)$ form themselves a Lax pair
\beq
\dot L(x)=[M(x),L(x)]\label{laxlmo}
\eeq
This is easily verified using (\ref{laxn}), since
\begin{eqnarray}
\dot L(x)&=&\sum_{k=0}^{N+1}\prod_{n=k+1}^{N+1}L_n(x)
\,\dot L_k\,\times \prod_{n=0}^{k-1}L_n(x)\\
&=&\sum_{k=0}^{N+1}\prod_{n=k+1}^{N+1}L_n(x)(M_{k+1}L_k-L_kM_k)
\prod_{n=0}^{k-1}L_n(x)\\
&=&\sum_{k=0}^{N+1}\prod_{n=k+1}^{N+1}L_n(x)\,M_{k+1}\,\prod_{n=0}^{k}
L_n(x)
-\sum_{k=0}^{N+1}\prod_{n=k}^{N+1}L_n(x)\,M_k\,\prod_{n=0}^{k-1}L_n(x)\\
&=& M_{N+2}L(x)-L(x)M_0(x).
\end{eqnarray}
In particular, the characteristic equation of $L(x)$ is time-independent
and defines a time-independent spectral curve
\beq
\G=\{(x,y); 0=R(x,y)\equiv {\rm det}\, (yI-L(x))\}.
\eeq
We assert that these spectral curves are
Landsteiner-Lopez curves (\ref{ll1}). In fact, it follows immediately
from the expression (\ref{laxodd}) that
${\rm det}\,L_n(x)=1$, $L_n(x)=L_n(x)^T$, and $L_n(x)^{-1}=L(-x)$. Thus
\beq
{\rm det}\,L(x)=1,\quad L(x)^{-1}=L(-x).
\eeq
These two equations imply that ${\rm det}(yI-L(x))$ is of the form
(\ref{ll1}) for some polynomial $f(x)$. To obtain the expression
(\ref{ll2}) for $f(x)$, it suffices to observe that
\beq
f(x)={\rm Tr}\, L(x)={\rm Tr}\,(1+x\sum_{n=0}^{N+1}s_ns_n^T)
+O(x^2)=3+O(x^2).
\eeq
Define the moduli $u_i$ of the curve $R(x,y)=0$ as in (\ref{ll1})
by $f(x)=3+x^2\sum_{i=0}^Nu_ix^i$. Then the correspondence between
the dynamical variables $s_n$, $0\leq n\leq N+1$, and the moduli $u_i$
is given by
\beq
u_i=\sum_{I_i}s_{n_1}^Ts_{n_2}s_{n_2}^Ts_{n_3}\cdots s_{n_{i-1}}^Ts_{n_i}
\label{2moduli}
\eeq
where the summation runs over the set $I_i$ of all ordered $i$-th
multi-indices $n_1<n_2<\cdots<n_i$.

To obtain the phase space of the model, we consider the
space of all $(N+2)$-periodic spin chains $s_n$,
subject to the constraint (\ref{con0}), and modulo the
equivalence $s_n\sim Vs_n$, where $V$ is a matrix satisfying
$V^TV=I$. The dimension of this space is
\beq
{\rm dim}\,\{s_n\}/\{s_n\sim Vs_n\} \ =\ 2N+1.\label{dimodd}
\eeq
Indeed, the $(N+2)$-periodic spin chains $s_n$ have $3(N+2)$
degrees of freedom. The constraint (\ref{con0}) removes $N+2$
degrees of freedom, and the equivalence $s_n\sim Vs_n$ removes
$3$ others, since the dimension of the matrices $V$ with $V^TV$ is
$3$. A $2N$-dimensional symplectic manifold ${\cal L}^{\rm odd}$
is obtained by setting
\beq
{\cal L}^{\rm odd}\ =\ \{s_n; \ u_N=constant\ \}/\{s_n\sim Vs_n\}
\eeq
On the space ${\cal L}^{\rm odd}$, the system is Hamiltonian
with respect to the symplectic form defined by the differential
$d\lambda_{(1)}=(\ln x){dy\over y}$, with Hamiltonian
\beq
H_{(1)}={u_{N-1}\over u_N}=\sum_{n=0}^{N+1}{(s_{n+1}^Ts_{n-1})\over
(s_{n+1}^Ts_{n})(s_n^Ts_{n-1})}
\eeq
The action-variables
are the periods of the differential
$d\lambda_{(1)}=-(\ln x){dy\over y}$ over a basis of $N$ cycles for
the curve $\Gamma$, which are odd under the involution $\sigma$.
If the curve $\Gamma$ is viewed as a two--sheeted cover of $\Gamma/\sigma$,
these $N$ odd curves can be realized as the $N$ cuts along which
the sheets are to be glued.

\subsection{The Even Divisor Spin Model}

The even divisor spin model is
the Hamiltonian reduction of a periodic spin chain
model which incorporates a natural gauge invariance.

\medskip

The starting point is a $(N+2)$-periodic chain of pairs
of three-dimensional complex
vectors $p_n=(p_{n,\a}),\ \ q_n=(q_{n,\a}),\ 1\leq\a\leq3$,
satisfying the constraints (\ref{con1}). We impose
the equations of motion (\ref{pq0}). As noted before,
the constraints and the equations of motion are invariant
under the gauge transformations (\ref{g1},\ref{g2}).
In particular, a gauge fixed version of the equations of motion
(\ref{pq0}) is
\beq\label{pq10}
\dot p_n={p_{n+1}\over p_{n+1}^Tq_n}+{p_{n-1}\over p_{n-1}^Tq_n},
\quad
\dot q_n=-{q_{n+1}\over p_n^Tq_{n+1}}-{q_{n-1}\over p_n^Tq_{n-1}}.
\eeq
This version follows from the other one by the gauge transformation
\beq
p_n\to \lambda_n(t)p_n,\ \  q_n\to \lambda_n^{-1}(t)q_n,\ \
\lambda(t)=\exp\left(-\int^t \mu_n(t')dt'\right).
\eeq
We shall see in the next section that the system (\ref{pq0}) admits
a Lax representation.

\medskip

A reduced system is defined as follows. We impose the additional
constraints (\ref{con}). With these constraints,
the spectral curves of the system are the Landsteiner-Lopez curves
(\ref{ll1}). The dimension of the phase space $\M$ of
all $(q_n,p_n)$ subjected to the previous constraints
and divided by the gauge group $G$ of (\ref{g1},\ref{g2}), is
\beq
{\rm dim}\,\M\equiv{\rm dim}\,\{(q_n,p_n)\}/G=2N.\label{dim}
\eeq
To see this, assume that $N$ is even (the counting for $N$
odd is similar). Then the constraint (\ref{con})
reduces the number of degrees of the $(N+2)$-periodic spin
chain $(q_n,p_n)$ to the number $3(N+2)$ of a $(N+2)$-periodic
spin chain. The constraint (\ref{con1}) and the gauge transformation
(\ref{g1}) each eliminates ${N\over 2}+1$
degrees of freedom. Now the dimension of the space of
matrices $W$ satisfying $Wg_0=g_0W$ is $5$. However,
in the gauge transformation (\ref{g2}), the matrices $W$ which are
diagonal have already been accounted for in the gauge transformation
(\ref{g1}). Altogether,
we arrive at the count which we announced earlier.

The phase space $\{(q_n,p_n)\}/G$ itself can be reduced further,
to a lower-dimensional phase space defined by suitable
constraints on the moduli space $(u_0,\cdots,u_N)$.
It turns out that there are 2 possible natural further reductions,
each related to its own choice of differential $d\lambda$
and corresponding Hamiltonian structure:

\medskip

$\bullet$ On the $(2N-2)$-dimensional phase space defined by the constraints
\beq
{\cal M}_0=\{(q_n,p_n); \ u_N=1,\ u_{N-1}=0\}/G
\eeq
the system is Hamiltonian with respect to the symplectic form defined
by the differential $d\lambda=x{dy\over y}$.
Here we have used the same notation for the space just introduced and
the space $\M_0$ described in the Main Theorem,
in anticipation of their isomorphism which will be established
later in \S 4. The Hamiltonian
is given by
$H=u_{N-2}$
or equivalently by (\ref{H12}).

The action-variables
are periods of $d\lambda$ along a basis of $N-1$
cycles $A_i$ of $\Gamma$ which are even under the involution
$\sigma$. (Equivalently, the $A_i$ correspond to a basis
of cycles for the factor curve $\Gamma/\sigma$.) This is the desired
integrable Hamiltonian system, corresponding to the $\N=2$ supersymmetric
$SU(N)$ Yang-Mills theory with a hypermultiplet in the anti-symmetric
representation.

\medskip

$\bullet$ On the $(2N-2)$-dimensional phase space ${\cal M}_2$
defined by the constraints
\beq
{\cal M}_2=\{(q_n,p_n);\ u_0={\rm constant},\ u_1={\rm constant}\}/G
\eeq
the system is Hamiltonian with respect to the symplectic form defined
by the differential $d\lambda_{(2)}=-{1\over x}{dy\over y}$.
This symplectic form coincides with the natural form
\beq
\omega=\sum_n dp_n^T\wedge dq_n
\eeq
with respect to which the system (\ref{pq0}) is manifestly
Hamiltonian, with Hamiltonian
\beq
H(p,q)=\ln u_N
={1\over 2}\sum_{n=0}^{N+1}\ln\left[(p_n^+q_{n-1})(p_{n-1}^+q_n)\right]
\eeq
The action-variables are the periods of the differential
$d\lambda_{(2)}=-{dy\over xy}$
over again the even cycles $A_i$ of the earlier case.

\section{The Direct and Inverse
Spectral Transforms}

We concentrate now on the even divisor spin model.
The main goal of this section is to
describe the map stated in the main theorem,
which associates to the spin chain $(q_n,p_n)$
a geometric data $(\Gamma, [D])$
\beq
(q_n,p_n)\to (\Gamma,[D]),\label{map}
\eeq
The curve $\Gamma$ is obtained by showing that the dynamical
system (\ref{pq0}) for $(p_n,q_n)$ admits a Lax representation
$\dot L(x)=[M(x),L(x)]$, in which case $\Gamma$ is the
spectral curve $\{\det\,(yI-L(x))=0\}$. The Lax operator $L(x)$
also gives rise to the Bloch function, which is
essentially its eigenvector. The divisor $D$ is obtained
by taking the divisor of poles of the Bloch function.
A characteristic feature of the even divisor spin model is
that the equivalence class of this divisor $[D]$ is {\it even}
under the involution $\sigma$.
The map (\ref{map})
descends to a map from the space of equivalence classes of $(q_n,p_n)$
under the gauge group $G$ to the space of geometric data $(\Gamma,[D])$.
These two spaces are of the same dimension $2N$: we saw this in (\ref{dim})
for the first space, while for the second, the number $2N$ of parameters
is due to $N+1$ parameters for the
Landsteiner-Lopez curves (including $u_N$ and $u_{N-1}$), and $N-1$
parameters for the even divisors $[D]$.
It is a fundamental fact in the theory
that the map (\ref{map}) becomes then a bijective correspondence
of generic points
\beq \label{map1}
\{q_n,p_n)\}/G \leftrightarrow \{(\Gamma,[D])\}
\eeq
We shall refer to the construction $\rightarrow$ described above as {\it the
direct problem}. The reverse construction $\leftarrow$,
which recaptures the dynamical variables $(p_n,q_n)$
from the geometric data $(\Gamma,[D])$ will
be referred to as {\it the inverse problem}.
As usual in the geometric theory of solitons \cite{k2},
it will be based on the construction of a
Baker-Akhiezer function. We now provide the details.

\subsection{The Lax Representation}

We exhibit first the Lax representation for the system (\ref{pq0}).
The desired formulas can be obtained from a slight modification
of the easier odd spin model treated in \S 3.1.
Let $p_n, q_n$ be $(N+2)$-periodic,
three-dimensional vectors satisfying $p_n^Tq_n=0$,
and define matrix-valued
functions $L_n(x)$ and $M_n(x)$ by
\beq
L_n(x)=1+x\,q_np^T_n,\ \ M_n(x)=x\left({q_{n-1}p^T_n\over p_n^Tq_{n-1}}-
{q_{n}p^T_{n-1}\over p_{n-1}^Tq_{n}}\right) \ .
\label{LM}
\eeq
Then a direct calculation shows that the matrix functions $L_n(x)$ and
$M_n(x)$ satisfy the Lax equation
\beq\label{laxLM}
\p_tL_n=M_{n+1}L_n-L_nM_n
\eeq
if and only if the vectors $p_n$ and $q_n$ satisfy the equations of
motion (\ref{pq0}).

As before, the equation (\ref{laxLM}) is a compatibility condition for the
linear system $\psi_{n+1}=L_n(x)\psi_n$,
$\dot\psi_n=M_n(x)\psi_n$. To obtain the spectral curve $\Gamma$,
we observe that the same arguments as in the case of the odd spin model
show that the matrix $M(x)=M_0(x)$ and the monodromy
matrix $L(x)$ defined by
$L(x)=\prod_{n=0}^{N+1}L_n(x)$
form again a Lax pair
\beq
\dot L(x)=[M(x),L(x)], \label{laxT}
\eeq
Thus the spectral curve $\Gamma=
\{(x,y);R(x,y)\equiv\det(yI-L(x))=0\}$
is time-independent and well-defined. We have used here the same notation $R(x,y)$
as for (\ref{ll1}), since the equation
$\det\,(yI-L(x))$ is indeed of the Landsteiner-Lopez form.
To see this, we note that ${\rm det}\,L_n(x)=1$ and
$L_n(-x)=L_n(x)^{-1}$. Together with the constraint (\ref{con}),
this implies
\beq
\det L(x)=1,\ \ L(-x)=g_0L^{-1}(x)g_0. \label{c2}
\eeq
But we also have near $x=0$
\beq
{\rm Tr} \ L(x) ={\rm Tr}\, (1+x\sum_{n=0}^N q_np_n^T)+O(x^2)=
3+0(x^2),
\eeq
so that $\det(yI-L(x))$ is of the form (\ref{ll1}).

\medskip

We observe that the expression $R(x,y)=\det\,(yI-L(x))$
is invariant with respect to the gauge transformations
(\ref{g1}) and (\ref{g2}). Therefore, if we write $R(x,y)$
in the Landsteiner-Lopez form (\ref{ll1})
with moduli $u_i$, the moduli $u_i$ are well-defined
functions on the factor-space $\M$. In analogy
with the odd spin case, $u_i$ can be written
in terms of the dynamical variables $(p_n,q_n)$ as
\beq\label{ui}
u_k=\sum_{I_k}(p_{i_1}^+q_{i_2})(p_{i_2}^+q_{i_3})\cdots(p_{i_k}^+q_{i_1})
\eeq
Here the summation is again over sets $I_k$ of multi-indices
$I=(i_1<i_2<\ldots<i_k)$.

\subsection{General Properties of Bloch Functions}

The points $Q=(x,y)$ of the spectral curve $\Gamma=\{(x,y);{\rm det}(yI-L(x))=0\}$
parametrize the Bloch functions $\{\psi_n(Q)\}_{0\leq n\leq N+1}$
of the spin model. We begin by recalling the definition of Bloch functions,
and by describing their main properties in the case of our model.

\medskip

$\bullet$ We fix a generic choice of moduli parameters $u_i$.
Then the matrix $L(x)$ has 3 distinct eigenvalues $y$,
except possibly at a finite number of points $x$. Let $Q=(x,y)$.
The Bloch solution $\psi_n(Q)$ for the spin model
$\{L_n(x)\}_{0\leq n\leq N+1}$ is the function $\psi_n(Q)$
with the following properties
\beq\label{bloch}
\Psi_{n+1}(Q)=L_n(x)\Psi_n(Q),\ \ \Psi_{N+n+2}(Q)=y\Psi_n(Q).
\eeq
These equations determine $\psi_n(Q)$ only up
to a multiplicative constant.
To normalize $\psi_n(Q)$, we observe that for generic moduli
parameters $u_i$, there are only finitely many points
$Q$ where the eigenvector
$\psi_0(Q)$ of the matrix $L(x)$ satisfies the linear
constraint $\sum_{\alpha=1}^3\psi_0^{\alpha}(Q)=0$.
Outside of these points, we can fix $\psi_n(Q)$
by the following normalization condition
\beq
\sum_{\a=1}^3\psi_0^{\a}=1. \label{nor}
\eeq
The Bloch function $\psi_n(Q)$ is then determined
on the spectral curve $\Gamma$
outside of a finite number of points, and hence uniquely
on $\Gamma$. Furthermore, the components of $\psi_n(Q)$
are meromorphic functions on $\Gamma$. This follows from
the constraint (\ref{nor}) and the equation $L(x)\psi_0(Q)=y\psi_0(Q)$.
They imply that $\psi_0(Q)$ is a rational expression in $y$ and in the entries
of the matrix $(L_{\a\b}(x)-L_{\a 3}(x))_{1\leq\a\leq 3\atop
1\leq\beta\leq 2}$, in view of Cramer's rule
for solving inhomogeneous systems of linear equations.
Since $x,y$ and $L_{\a\b}(x)$ are all
meromorphic functions on $\Gamma$, our assertion follows.

$\bullet$ The exceptional points excluded in the preceding construction
of Bloch functions are the points where $L(x)$ has multiple eigenvalues,
and the points where the eigenvector $\psi_0(Q)$ lies in the linear
subspace of equation $\sum_{\a=1}^3\psi_0^\a(Q)=0$.
By restricting ourselves to generic values of the moduli $u_i$,
we can make the convenient
assumption that these two sets of points
are disjoint. In this case, it is evident that
at points where $\sum_{\a=1}^3\psi_0^\a(Q)=0$, the function
$\psi_0(Q)$ develops a pole.

Consider now a point $x_0\not=0$ where the matrix $L(x)$
has a multiple eigenvalue. Let $(x-x_0)^{1/b}$ be the
local holomorphic coordinate centered at the points $Q$
lying above $x_0$, where the branching index $b$
can be either $1$ or $2$. (We can exclude the possibility $b=3$
by a genericity assumption on the moduli $u_i$.) The holomorphic
function $y$ on the
surface $\Gamma$ can be expanded as
\beq
y=y_0+\epsilon y_1(x-x_0)^{1/b}+O(x-x_0),\label{puiseux}
\eeq
where $\epsilon^b=1$ is a root of unity.
If $b=1$, it follows that
\beq
\partial_x R(x_0,y_0)=\partial_yR(x_0,y_0)=0,
\eeq
which means that the curve is singular at $(x_0,y_0)$.
By a genericity assumption on the moduli $u_i$, the only
singular point on $\Gamma$ is at $x_0=0$, and this possibility
has been excluded. Thus $b=2$, and the curve $\Gamma$ has a branch point
at $x_0$ if and only if $L(x_0)$ has multiple eigenvalues.
The matrix $L(x_0)$ can now be shown to be a Jordan cell, i.e.,
$L(x_0)$ is of the form
\beq
L(x_0)=\left(\begin{array}{ccc}
 \lambda_1 & \mu & 0 \\
0& \lambda_2 &0\\
0& 0 & \lambda_3
\end{array}\right)\label{jordan}
\eeq
in a suitable basis, for some $\mu\not=0$ and $\lambda_1=\lambda_2\not=\lambda_3$.
In fact, $L(x_0)$ has only one double eigenvalue
by genericity assumptions on $u_i$. The three branches
of the function $y$ consist then of one branch which is
of the form $\lambda_3+y_1(x-x_0)+\cdots$
and is holomorphic
in the variable $x-x_0$. The other two branches
are of the form
\beq\label{ypm}
y=\lambda_1\pm y_1(x-x_0)^{1/2}+\cdots.
\eeq
We must have $y_1\not=0$,
for otherwise $y=O(x-x_0)$, and the same argument
which ruled out the branching index $b=1$ would imply
that $\Gamma$ is singular at $x_0$.
Now for $x$ near but distinct from $x_0$,
the Bloch function $\psi_0(Q)$ also has 3 distinct branches.
Let $\psi_{\pm}$ be the branches corresponding to the eigenvalues
in (\ref{ypm}), and expand them as
\beq
\psi_{\pm}=\psi_{\pm}^{(0)}+(x-x_0)^{1/2}\psi_{\pm}^{(1)}+O(x-x_0)
\label{puiseux2}
\eeq
Up to $O(x-x_0)$, the eigenvector condition can be expressed as
\beq
L(x)(\psi_{\pm}^{(0)}+(x-x_0)^{1/2}\psi_{\pm}^{(1)})
=(\lambda_1\pm y_1(x-x_0)^{1/2})(\psi_{\pm}^{(0)}+(x-
x_0)^{1/2}\psi_{\pm}^{(1)})
\eeq
This is equivalent to
\beq
L(x_0)\psi_{\pm}^{(0)}=\lambda_1\psi_{\pm}^{(0)},
\quad\quad
(L(x_0)-\lambda_1)\psi_{\pm}^{(1)}=\pm y_1\psi_{\pm}^{(0)}.
\eeq
Clearly, this equation admits no solution if $L(x_0)$ is diagonal.
Thus $L(x_0)$ is of the form (\ref{jordan}) with $\mu\not=0$.
We can now identify the coefficients $\psi_{\pm}^{(0)}$ and
$\psi_{\pm}^{(1)}$ in the Puiseux expansion (\ref{puiseux2}).
The eigenspace of $L(x_0)$ corresponding to
the eigenvalue $\lambda_1$ is one-dimensional and generated
by a single vector $\phi_1$,
which we can take to satisfy the normalization
condition (\ref{nor}). Evidently,
$\psi_{\pm}^{(0)}=\phi_1$.
Let $\phi_2$
be the second basis vector in the basis with respect to
which $L(x_0)$ takes the Jordan form (\ref{jordan}), i.e.,
$L(x_0)\phi_2=y_1\phi_2+\mu\phi_1$.
Then the second equation above is solved by
\beq
\psi_{\pm}^{(1)}=\pm ({y_1\over\mu}\phi_2+\nu\phi_1),
\eeq
where the constant $\nu$ is chosen so that $\sum_{\a=1}^3
\psi_{\pm}^{(1),\a}=0$.

$\bullet$ Outside a finite number of points $x$, the matrix $L(x)$
has 3 distinct eigenvalues $y(a)$ and three distinct eigenfunctions
$\psi_0(a)$, $1\leq a\leq 3$, normalized uniquely by the condition
(\ref{nor}). The function
\beq
{\rm det}^2\big\{\psi_0(1)\ \psi_0(2)\ \psi_0(3)\big\}
\label{det}
\eeq
is independent of the ordering of both
$\psi_0(a)$ and the corresponding eigenvalues $y(a)$.
By the preceding observations, it can be expressed
as a rational function of $x$ and $y(a)$, which is also
symmetric under permutations of $y(a)$. Thus it is actually
an unambiguous and rational function of $x$.
We observe that the function
${\rm det}^2\big\{\psi_0(1)\ \psi_0(2)\ \psi_0(3)\big\}$
vanishes at exactly those values of $x$ which are
branch points for the spectral curve ${\rm det}(yI-L(x))=0$.
Indeed, we saw earlier that the branch points $x_0$ are exactly
the points where $L(x_0)$ has multiple eigenvalues.
Outside points $x_0$ where $L(x_0)$ has multiple eigenvalues,
the determinant (\ref{det}) is readily seen to be $\not=0$
(it may be infinite, because of the normalization (\ref{nor})).
Conversely, assume that $x_0$ is a branch point. Then our preceding
discussion shows that for $x$ near $x_0$
\beq
{\rm det}^2\big\{\psi_0(1)\ \psi_0(2)\ \psi_0(3)\big\}(x)
=
(x-x_0){\rm det}^2\big\{\phi_1\ \phi_2\ \psi_0(3)\big\}
+
O(x-x_0)^{3/2}
\eeq
This shows that ${\rm det}^2\big\{\psi_0(1)\ \psi_0(2)\ \psi_0(3)\big\}(x_0)
={\rm lim}_{x\to x_0}{\rm det}^2\big\{\psi_0(1)\ \psi_0(2)\ \psi_0(3)\big\}(x)
=0$, establishing the observation. Furthermore, since the
vectors $\phi_1$, $\phi_2$, and $\psi_0(3)$ are linearly
independent by construction, we obtain the
important fact that the order of vanishing of the
square of the determinant in (\ref{det}) at a branch point
is exactly $1$. (More generally, for an arbitrary
branching index $b$, the order of vanishing of the
square of the determinant is equal to $b-1$, although
we do not need this more general version here, thanks
to our genericity assumption on the moduli $u_i$.)

$\bullet$ We can now determine the number of poles of the Bloch
function $\psi_0(Q)$ outside of the points $P_a$ above $x=\infty$.
Clearly, this number is half of the number of poles of the
expression (\ref{det}) outside of $x=\infty$.
Now at $x=\infty$, we saw that the operator $L(x)$ has 3
eigenvalues, so that (\ref{det}) does not vanish there.
Furthermore, we shall show later
that $\psi_0(Q)$ is finite at all three points
above $x=\infty$. Thus (\ref{det}) has neither a zero
nor a pole at $x=\infty$.
In view of the preceding discussion, the number of zeroes
of (\ref{det}) is equal to the number of branch points of
$\Gamma$. We showed earlier, using the Riemann-Hurwitz formula,
that the number of branch points of $\Gamma$ is $4N+2$.
It follows that the number of poles, and hence of zeroes
of $\psi_0(Q)$ on $\Gamma$ is $2N+1$.

$\bullet$ The poles of $\psi_n(Q)$ outside of the
points $P_a$ lying above $x=\infty$ are independent of $n$.
To see this, we let $S_0(x)$ be the $3\times 3$ identity matrix
$I$, and set $S_{n}(x)=L_{n-1}(x)S_{n-1}(x)=\prod_{0\leq k\leq n-1}L_k(x)$.
Then $\psi_n(Q)$ can be expressed as
\beq
\psi_{n}(Q)=L_{n-1}(x)\psi_{n-1}(Q)
=\prod_{0\leq k\leq n-1}L_k(x)\psi_0(Q)
=
S_n(x)\psi_0(x)
\eeq
This shows that the poles of $\psi_n(Q)$ outside of $P_a$
can only occur at the poles of $\psi_0(Q)$.
For generic values of the moduli $u_i$, we can assume that
all the poles of $\psi_n(Q)$, $0\leq n\leq N+1$,
are exactly of the same order $1$ when they occur outside of the
points $P_a$ above $x=\infty$.

\medskip

$\bullet$ Let $D=\{z_1,\cdots,z_{2N+1}\}$ be the divisor of poles
of the Bloch function $\psi_n(Q)$. Then a fundamental property of
the even divisor spin chain model is the invariance of the equivalence
divisor class $[D]$ of $D$ under the involution $\sigma$
\beq
[D]=[D^{\sigma}]\label{evendiv}
\eeq
In other words, there exists a meromorphic function on $\Gamma$
with poles at $z_n$ and $z_n^{\sigma}$. This is a consequence of how
$L(x)$ transforms under the involution $x\to-x$, $y\to y^{-1}$
\beq
L(-x)=g_0L(x)^{-1}g_0.
\eeq
This transformation rule implies that $g_0\psi_0(Q)$ is a
Bloch function at $(-x,y^{-1})$
if $\psi_0(Q)$ is a Bloch function at $(x,y)$. Thus $g_0\psi_0(Q)$
must coincide with $\psi_0(Q^{\sigma})$ up to normalization
\beq
g_0\psi_0(Q)=f(Q)\psi_0(Q^{\sigma})
\eeq
Since both $\psi_0(Q)$ and $\psi_0(Q^{\sigma})$ are meromorphic functions,
the function $f(Q)$ is meromorphic. This proves (\ref{evendiv}).

\medskip

We summarize the discussion in the following lemma:

\begin{lem}
The vector-function $\psi_n(Q)$ is a meromorphic vector-function on $\G$.
Outside the punctures $P_{a}$ (which are the points of $\G$ situated over
$x=\infty$) it has $g+2=2N+1$ poles $\{z_1,\ldots,z_{2N+1}\}$, which
are {\it $n$-independent}. The divisor class $[D]$ of $D$ is invariant
with respect to the involution $\s$, i.e.
there exists a function $f(Q)$ on $\G$ with poles at $z_j$ and zeroes at
$z_j^{\s}$.
\end{lem}

\subsection{The Direct Problem}

In the previous discussion, we made use only of the fact that
the curve $R(x,y)=0$ is the spectral curve of a matrix $L(x)$
which satisfies the involution condition $L(-x)=L(x)^{-1}$.
In particular, the discussion applies for generic values of the moduli
$u_i$ parametrizing the curves.

We consider now the direct problem for the system (\ref{pq0}),
where the matrix $L(x)$ arises more specifically
in terms of the dynamical variables $(q_n,p_n)$ as
$L(x)=\prod_{n=0}^{N+1}L_n(x)=\prod_{n=0}^{N+1}(1+xq_np_n^T)$.
The discussion in the previous section has provided a precise
description of the right hand side of the map (\ref{map}).
It is also evident that the map descends to the equivalence
classes of $(q_n,p_n)$ under the gauge group $G$.

It is convenient to exploit the gauge transformation (\ref{g2})
to normalize the Bloch functions at $x=0$.
First, we observe
that $L_n(0)=I$ for all $n$, so that $\psi_n(Q_a)$ is independent of $n$.
Furthermore, the Lax operator $L(x)$ can be written near $x=0$ as
\beq
L(x)=I+xT+O(x^2)
\eeq
where the matrix $T$ is given by
\beq
T=\sum_{n=0}^{N+1}
q_np_n^T
\eeq
In particular, $T$ satisfies the condition
\beq
T=g_0Tg_0.\label{newT}
\eeq
in view of the constraint $p_n=g_0p_{-n-1}$, $q_n=g_0q_{-n-1}$.
Next, recall from our discussion of the Landsteiner-Lopez curve in \S 2
that $T$ has 3 distinct eigenvalues $y_1(Q_{a})$, and that $y$
can be expanded as $y=1+y_1(Q_{\a})x+O(x^2)$ near $Q_a$.
Expanding $\psi_0(Q)$ near $Q_a$ as
$\psi_0(Q)=\psi_0(Q_a)+O(x)$, and using the preceding expansion for
$L(x)$, the condition $L(x)\psi_0(Q)
=y\psi_0(Q)$ for Bloch functions can be rewritten as
\beq
(I+xT)(\psi_0(Q_a)+x\psi_0'(Q_a))
=
(1+y_ax)(\psi_0(Q_a)+x\psi_0'(Q_a))
+O(x^2)
\eeq
This implies
\beq
T\psi_0(Q_a)=y_a\psi_0(Q_a)
\eeq
i.e., $\psi_0(Q_a)$ are precisely the three
eigenvectors of $T$, corresponding to the eigenvalues $y_a$.
If we let $\Psi_0(0)$ be the $3\times 3$ matrix whose columns
are the vectors $\psi_0(Q_a)$, then the transformation law (\ref{newT})
implies that $\Psi_0(0)$ satisfies the condition
\beq
\Psi_0(0)=g_0\Psi_0(0)g_0.
\eeq
Now the transformation (\ref{g2}) on $(q_n,p_n)$ does not change the
curve $\G$ and the divisor $D$, but changes the matrix $\Psi_0(0)$
into $W\Psi_0(0)$. But $\Psi_0(0)$ commutes with the matrix $g_0$,
and hence so does its inverse. This means that the inverse qualifies as
one of the gauge transformations $W$ allowed in (\ref{g2}).
Under such a gauge transformation $W$, the Bloch function $\Psi_0(0)$
gets transformed to the identity
\beq
\Psi_0(0)=I.\label{nor1}
\eeq
Henceforth we can assume then this normalization, and $p_n,q_n$ satisfies
the condition
\beq \label{nor2}
T_{\a}^{\b}=\sum_{n=0}^{N+1}q_{n,\a}p_n^{\b}=y_1^{\a}\delta_{\a}^{\b}.
\eeq

Our main task is to establish that the map (\ref{map1})
is generically locally invertible. This is the goal
of the next section on the inverse spectral problem,
but in order to motivate the constructions given there,
we identify here the basic
behavior of the Bloch function $\psi_n(x,y)$
near the points $P_{\a}$ above $x=\infty$.
For $(x,y)$ near $P_{\a}$, set
\beq
\psi_n(x,y)
=x^{p_{n\a}}\sum_{k=0}^{\infty}\psi_{n,k}(P_{\a})x^{-k}\label{expansion}
\eeq
Here $p_{n\a}$ is the order of the pole (or zero when $p_{n\a}<0$)
of $\psi_n(x,y)$ near $P_{\a}$, which may vary with both $n$ and $\a$.
The following lemma identifies the coefficients $\psi_{n,k}(P_{\a})$
up to normalization:

\begin{lem}
$\bullet$ In the neighborhood of the puncture $P_1$ (where $y=O(x^{N+2})$),
the vector-function $\psi_n$ has a pole of order
$n$ and the leading coefficient $\psi_{n,0}(P_1)$ of its expansion
is equal to
\beq
\psi_{n,0}(P_1)=\a_nq_{n-1}, \label{e2}
\eeq
where the scalar $\a_n$ satisfy the recurrence relation
\beq
\a_{n+1}=(p_n^Tq_{n-1})\a_n. \label{e3}
\eeq
The next coefficient $\psi_{n,1}(P_1)$ satisfies
\beq
\psi_{n+1,1}=\psi_{n,0}+q_n(p_n^T\psi_{n,1})\label{e4}
\eeq
$\bullet$ In the neighborhood of the puncture $P_3$ (where $y=O(x^{-N-2})$)
the vector-function $\psi_n$ has a zero of order
$n$ and the leading coefficient $\psi_{n,0}(P_3)$ of its expansion
is equal to
\beq
\psi_{n,0}(P_3)=\b_nq_n, \label{e5}
\eeq
where the scalar $\b_n$ satisfies the recurrence relation
\beq
\b_{n+1}=-{1\over(p_n^Tq_{n+1})}\b_n. \label{e6}
\eeq
$\bullet$ In the neighborhood of the puncture $P_2$ (where $y=1$)
the vector-function $\psi_n$ is regular and its evaluation $\psi_{n,0}(P_2)$
at $P_2$ is orthogonal to both $p_n$ and $p_{n-1}$, i.e.,
\beq
p_n^T\psi_{n,0}(P_2)=p_{n-1}^T\psi_{n,0}(P_2)=0.
\label{e7}
\eeq
\end{lem}

\noindent{\it Proof.}
First, we show that for generic moduli,
the Bloch function $\psi_0(x,y)$ is regular near each $P_{\a}$.
Observe that $\psi_{N+2}(x,y)=L(x)\psi_0(x,y)=y\psi_0(x,y)$.
Now the relation $\psi_{n+1}=L_n(x)\psi_n$ can be inverted to
produce
\beq
\psi_n(x,y)=L_n(x)^{-1}\psi_{n+1}(x,y)=(1-xq_np_n^T)\psi_{n+1}(x,y)
\eeq
Applying this relation $N+2$ times, we may write
\beq
\psi_0(x,y)=y(1-xq_0p_0^T)\cdots(1-xq_{N+1}p_{N+1}^T)\psi_0(x,y)
\eeq
Consider first the neighborhood of the point $P_3$, where $y$ is of
order $x^{-(N+2)}$. If $\psi_0(x,y)$ admits the expansion (\ref{expansion})
near $P_3$ with $\psi_{0,0}(P_3)\not=0$, then we must have
\beq
\psi_{0,0}=(-)^{N+2}q_0(p_0^Tq_1)\cdots(p_N^Tq_{N+1})(p_{N+1}^T\psi_{0,0})
\eeq
This shows that $\psi_{0,0}(P_3)$ is proportional to the vector
$q_0$, say $\psi_{0,0}=\b q_0$. Now recall that the Bloch function $\psi_0(x,y)$
satisfies
the normalization condition (\ref{nor}) throughout. This implies that
$\sum_{\a=1}^3\psi_{0,0}(P_3)=0$ if the order $n_0(P_3)$ of the pole
of $\psi_0(x,y)$ at $P_3$ is positive. For generic values of the moduli
of the curve $\Gamma$, we may assume that $\sum_{\a=1}^3q_{0\a}\not=0$.
It follows that $\b_0=0$ and hence $\psi_{0,0}(P_3)=0$,
which contradicts the definition of $\psi_{0,0}(P_3)$.
This shows that $n_0(P_3)=0$, and the Bloch function $\psi_0(x,y)$
is regular at $P_3$. The arguments near $P_1$ is similar and even more
direct, just using the equation $y\psi_0(x,y)=\psi_{N+2}(x,y)
=\prod_{n=0}^{N+1}(1+xq_np_n^T)\psi_0(x,y)$.
It shows, incidentally, that the leading coefficient $\psi_{0,0}(P_1)$
is proportional to $q_{N+1}$.
At $P_2$, the regularity of $\psi_0(x,y)$
follows from the regularity of $\psi_0(x,y)$ at the
other two points $P_1$ and $P_3$, and from the fact that for generic moduli,
the determinant (\ref{det}) is regular.

It is now easy to see that the functions $\psi_n(x,y)$ have the zeroes
and poles spelled out in Lemma 4.2. The recurrence relations stated there
can also be read off the defining relations $\psi_{n+1}=L_n(x)\psi_n(x)$.
For example, near $P_1$, we find
\beq
x^{n+1}\left(\psi_{n+1,0}+{1\over x}\psi_{n+1,1}+\cdots\right)
=x^n(1+xq_np_n^T)\left(\psi_{n,0}+{1\over x}\psi_{n,1}+\cdots\right)
\eeq
This implies
\begin{eqnarray}
\psi_{n+1,0}&=&q_n (p_n^T\psi_{n,0})\\
\psi_{n+1,1}&=&\psi_{n,0}+q_n(p_n^T\psi_{n,1})
\end{eqnarray}
The relations (\ref{e3}, \ref{e4}) follow. Near $P_2$, we write instead
\beq
x^{-n}\left(\psi_{n,0}+{1\over x}\psi_{n,1}+\cdots\right)
=x^{-n-1}(1-xq_np_n^T)\left(\psi_{n+1,0}+{1\over x}\psi_{n+1,1}+\cdots\right)
\eeq
This implies
\begin{eqnarray}
\psi_{n,0}&=&-q_n (p_n^T\psi_{n+1,0})\\
\psi_{n,1}&=&\psi_{n+1,0}-q_n(p_n^T\psi_{n+1,1})
\end{eqnarray}
which gives (\ref{e5}, \ref{e6}). Finally near $P_2$, we get
\beq
\psi_{n+1,0}+{1\over x}\psi_{n+1,1}+\cdots
=\left(1+xq_np_n^T)(\psi_{n,0}+{1\over x}\psi_{n,1}+\cdots\right)
\eeq
This implies that $p_n^T\psi_{n,0}=0$. Furthermore,
$\psi_{n+1,0}= \psi_{n,0}+q_n(p_n^T\psi_{n,1})$. Multiplying
on the left by $p_n^T$, we conclude that $p_n^T\psi_{n+1,0}=0$.
This establishes (\ref{e7}), and Lemma 4.2 is proved.

\subsection{The Inverse Spectral Problem}

It is now a standard procedure in the geometric
theory of soliton equations to solve the inverse problem
using the concept of the Baker-Akhiezer function originally proposed in
\cite{k2}. The main properties of the Baker-Akhiezer function
in our model are the following.

\medskip

$\bullet$ Let $\G$ be a Landsteiner-Lopez curve defined by equation (\ref{ll1}).
Then for a divisor $D$ of degree $g+2=2N+1$ in general position,
there exists a unique vector-function $\phi_n(t,Q)$ such that:

(a) $\phi_n(t,Q)$ is meromorphic on $\G$ outside the punctures $P_1,P_3$.
It has at most simple poles at the points $z_i$ of the divisor $D$
(if all of them are distinct);

(b) In the neighborhood of the punctures $P_1$ and $P_3$,
it has respectively the form
\beq
\phi_n=x^ne^{xt}\left(\sum_{k=0}^{\infty}\phi_{n,k}(P_1)x^{-k}\right),\ \ Q\to P_1,
\label{ex1}
\eeq
\beq
\phi_n=x^{-n}e^{-xt}\left(\sum_{k=0}^{\infty}\phi_{n,k}(P_3)x^{-k}\right),\ \ Q\to
P_3,
\label{ex11}
\eeq

(c) At the points $Q_{a}$, $\phi_n(Q)$ is regular,
and $\phi_n(Q_{a})$ is equal to
\beq\label{n10}
\phi_{n,\a}(t,Q_{\b})=\delta_{\a,\b}.
\eeq

The arguments establishing the existence of the Baker-Akhiezer
function $\phi_n$ are well-known, so we shall be brief.
First, we recall that as shown in
\cite{k2} for any algebraic curve with two punctures, any fixed local
coordinate in the respective neighborhoods of the punctures,
and for any divisor $D$ of degree $g$
there exists a unique (up to a constant factor) function with
the analytic properties stated above.
Now let $(P_1,P_3)$ be the punctures, and
let $x^{-1}$ be the local coordinate near either one of
the punctures. We can easily
show that if $D$ has degree $g+2$, the dimension of the space of such
functions is equal to $3$. We form the 3-dimensional vector whose components
are just the three independent functions from this space. This 3-dimensional
vector is unique
up to multiplication by a constant matrix. We fix this matrix by
the normalization condition (\ref{n10}). This establishes our claim.

The function $\phi_n(t,Q)$ can be written explicitly in
terms of the Riemann $\theta$-function associated with $\Gamma$. The
$\theta$-function is
an entire function of $g=2N-1$ complex variables
$z=(z_1,\dots,z_g)$, and is defined by its Fourier expansion
$$
\theta(z_1,\dots,z_g)=\sum\nolimits_{m\in {\bf Z}^g}
e^{2\pi i<m,z>+\pi i <\tau m,m>},
$$
where $\tau=\tau_{ij}$ is the period matrix of $\G$.
The $\theta$-function has the following monodromy properties with respect to
the lattice ${\bf Z}^g+\tau{\bf Z}^g$
$$
\theta(z+l)=\theta(z),\qquad
\theta(z+\tau l)=\exp[-i\pi <\tau l,l>-2i\pi <l,z>]\,\theta(z)
$$
where $l$ is an integer vector,  $l\in {\bf Z}^g$.
The complex torus $J(\Gamma)={\bf C}^g/{\bf Z}^g+\tau {\bf Z}^g$ is the
Jacobian variety of the curve $\Gamma$. The Abel transform
$$
\G\ni Q\to A_k(Q)=\int_{Q_0}^Q d\omega_k
$$
imbeds the curve $\Gamma$ into its Jacobian variety. Here $d\omega_k$
is a basis of $g$ holomorphic differentials, normalized as dual to
the $A$-cycles of a symplectic homology basis for $\G$.

According to the Riemann\,--\,Roch theorem, for each divisor
$D=z_1+\ldots+z_{g+2}$
in the general position, there exists a unique meromorphic function
$r_{\alpha}(Q)$ with
$r_{\alpha}(Q_{\beta})=\delta_{\alpha\beta}$
and $D$ as the divisor of its poles.
It can be written explicitly as
(see details in \cite{bab}):
$$
r_{\alpha}(Q)=\frac{f_{\alpha}(Q)}{f_{\alpha}(Q_{\alpha})},\qquad
f_{\alpha}(Q)=\theta(A(Q)+Z_{\alpha})
\frac{\prod_{\beta\neq \alpha} \theta(A(Q)+F_{\beta})}{%
\prod_{m=1}^l\theta (A(Q)+S_m)},
$$
where
$$
F_{\beta}={}-{\cal K}-A(Q_{\beta})-\sum_{j=1}^{g-1} A(z_j),\qquad
S_m={}-{\cal K}-A(z_{g-1+m})-\sum_{j=1}^{g-1} A(z_j),
$$
$$
Z_{\alpha}=Z_0-A(R_{\alpha}), \quad Z_0={}-{\cal K}-\sum_{j=1}^{g+2}
A(z_j)+\sum_{\alpha=1}^3 A(Q_{\alpha}),
$$
where ${\cal K}$ is the vector of Riemann constants.

Let $d\Omega_0$ and $d\Omega_1$ be the unique normalized
meromorphic differentials on $\Gamma$, which are holomorphic outside $P_1$
and $P_3$, and with the property that $d\Omega_0$ has simple poles at the punctures
with residues $\mp 1$, $d\Omega_1$ is regular at $P_3$, and has
the form $d\Omega_1=dx(1+O(x^{-1}))$ at $P_1$.
The normalization means that the differentials
have zero periods around $A$-cycles
$$
\oint_{A}d\Omega_0=\oint_{A}d\Omega_1=0.
$$
We observe that the differential $d\Omega_1^{\s}(Q)=d\Omega(Q^{\s})$ has a
pole only at $P_3$, and
is there of the form $-dx(1+O(x^{-1})$.

Let $V$ and $U$ be the vectors whose components
are the $B$-periods of the differentials $d\Omega_0$
and $d\Omega_1$ respectively
$$
V=\frac{1}{2\pi i} \oint_{B} d\Omega_1, \
U=\frac{1}{2\pi i} \oint_{B} d\Omega_0.
$$
The Baker-Akhiezer function $\phi_n(t,Q)$ is given by
\beq\label{BA}
\phi_{n,\a}(t,Q)=r_{\alpha}(Q)
\frac{\theta(A(Q)+tU^++nV+Z_{\alpha})\;\theta(Z_0)}{%
\theta(A(Q)+Z_{\alpha})\;\theta(tU^++nV+Z_0)}
\exp{\left(\int_{Q_{\alpha}}^Q nd\Omega_0+td\Omega^+\right)}
\eeq
where $d\Omega^+=d\Omega_1+d\Omega_1^{\s}$ and $U^+=U+U^{\s}$.
\medskip

$\bullet$
The Baker-Akhiezer function $\phi_n$ is a Bloch function,
in the sense that
\beq
\phi_{N+2+n}(t,Q)=y\phi_n(t,Q). \label{B}
\eeq
This is just a consequence of the fact that both sides of the equation
satisfy the criteria for the Baker-Akhiezer function, and that
the Baker-Akhiezer function is unique. Similarly, the uniqueness
of the Baker-Akhiezer function implies that,
if the divisor $D$ is equivalent to $D^{\s}$, then
the function $\phi_n$ satisfies
\beq
\phi_n(t,Q)=g_0\phi_{-n}(t,Q^{\s})f(Q), \label{sym}
\eeq
where $f(Q)$ is a function with poles at $\g_s$ and zeros at $\g_s^{\s}$.
Without loss of generality, we may assume that $f(Q_1)=f(Q_3)=-f(Q_2)=1$.

\medskip

$\bullet$
Let $\phi_n(t,Q)$ be the Baker-Akhiezer function corresponding
to $\G$ and the divisor $D$ of degree $2g+2$.
Let $p_n(t)$ be a vector orthogonal to $\phi_{n,0}(P_3,t)$ (the leading term
in the expansion (\ref{ex1})), and to $\phi_n(t,P_2)$, i.e.
\beq\label{f1}
p_n^T\phi_{n,0}(P_3)=p_n^T\phi_n(t,P_2)=0,
\eeq
and $q_n$ be the vector
\beq\label{f2}
q_n={\phi_{n,0}(P_1)\over p_n^T\phi_{n-1,0}(P_1)}.
\eeq
The vector functions $p_n,q_n$ are then $(N+2)$-periodic and mutually
orthogonal, and they satisfy the contraint (\ref{nor1}).
As can be expected from the gauge invariance (\ref{g1})
in the direct problem, the functions $p_n(t)$ and $q_n(t)$
which we obtain this way are defined only up
to a multiplier $\mu_n(t)$. However, the operators $L_n$ and $M_n(x)$
are uniquely defined by the expression (\ref{LM}).
Furthermore, again by uniqueness of the Baker-Akhiezer function,
the Baker-Akhiezer function $\phi_n(Q)$ satisfies
\beq\label{laxlm}
\psi_{n+1}(t,Q)=L_n(x)\psi_n(t,Q),\ \ (\p_t-M_n(x))\psi_n(t,Q)=0.
\eeq
Thus the vector function $(q_n(t),p_n(t)$ is a solution of
the dynamical system (\ref{pq0}).
If the equivalence class of the divisor $D$ is invariant with respect to
$\s$, then $(p_n,q_n)$ satisfies in addition the relation (\ref{con}).

\medskip

$\bullet$ The Baker-Akhiezer function $\phi_n(t,Q)$ satisfies the
same defining Bloch property (\ref{B}) as the Bloch function $\psi_n(Q)$,
except for the different normalizations, which is (\ref{n10})
in the case of $\phi_n(t,Q)$ and (\ref{nor}) in the case
of $\psi_n(Q)$. It follows that
\beq\label{10}
\psi_n(t,Q)=r^{-1}(t,Q) \phi_n(t,Q),\ \ r(t,Q)=\sum_{\a=1}^3 \phi_{0,\a}(Q)
\eeq
is a Bloch solution of (\ref{bloch}) normalized by the condition (\ref{nor}).
This leads to the following description of the dynamical system
(\ref{pq0}).

Let $p_n(t),q_n(t)$ be vector functions (subject to
constraints (\ref{con1},\ref{con},\ref{nor1}) )
which satisfy the equations (\ref{pq0}). Then the $t$-dependence
of the divisor $D$ under the map (\ref{map})
\beq
(p_n(t),q_n(t)) \longmapsto \{\G,D(t)=\sum_{j=1}^{2N+1}z_j(t)\}  \label{m10}
\eeq
coincides with the dynamics of the zeroes of
the function $r(t,Q)$ given by (\ref{10}).
The dynamics of the Bloch eigenfunction of (\ref{bloch}) (i.e. normalized
by (\ref{nor})) are described by
\beq\label{12}
(\p_t - M_n(t,x))\psi_n(t,Q)=\mu(t,Q)\psi_n(t,Q),\ \ \mu=-\p_t\ln r(t,Q)
\eeq
We observe that the linearization of the equations of motion on the Jacobian of
the curve is a direct corollary of the linear dependence on $t$ of the
exponential factor in the expansion of $\psi_n(t,Q)$ near the punctures.

\medskip

$\bullet$
As we saw earlier, the normalization (\ref{nor1}) can be achieved by the
action (\ref{g2}) of a subgroup of matrices $W$ which commutes with $g_0$.
In order to get $\M$, we have to consider in addition the action of
diagonal matrices $W$. The basic observation is the following.

Let the Baker-Akhiezer functions $\psi_n(t,Q)$ and $\psi'(t,Q)$
corresponds to equivalent divisors $D$ and $D'$, respectively.
Then
\beq
\psi_n'(t,Q)=W\psi_n(t,Q)h(Q), \label{eq}
\eeq
where $h(Q)$ is a function with poles at $D$ and zeros at $D'$, and
$W$ is a diagonal matrix $W=h^{-1}(Q_{\a})\delta_{\a,\b}$.
To establish (\ref{eq}), it suffices to check that both sides of
the equation have the same analytical properties.
The equation (\ref{eq}) implies that the vectors
$p_n,q_n$ defined by equivalent divisors
are related by a transformation (\ref{g2}) with a diagonal matrix $W$.
Altogether, we have established the following part of the
Main Theorem of \S 1:

\bigskip

\noindent
{\bf Theorem 1}.
{\it The map (\ref{map}) identifies the reduced phase space
$\M$ with a bundle over the space of algebraic curves $\G$ defined
by (\ref{ll1}) with $f_N(x)$ of the form (\ref{ll2}). At generic data,
the map has bijective differential. The fiber
of the bundle is the Jacobian $J(\G_0)$ of the factor-curve $\G_0=\G/\s$.}
\beq\label{m25}
\M=\{\G,[D]\in J(\G_0)\}
\eeq

\section{Hamiltonian theory and Seiberg-Witten differential: The Even
Divisor Model}

We come now to the crucial issue of how to determine the symplectic
forms with respect to which the system (\ref{pq0}) is Hamiltonian.
For this, we rely on the Hamiltonian approach proposed
in \cite{kp1} and \cite{kp2} for general
soliton equations expressible in terms of Lax or
Zakharov-Shabat equations. This approach was effective
in the study of gauge theories with matter in the fundamental
representation. Further applications were given in \cite{el} and \cite{k3}.
We review its main features.

\subsection{The Symplectic Forms in terms of the Lax Operator}

In order to find the Hamiltonian structure of the equations starting with
the Lax operator, we need to identify a two-form
on the phase space $\M$ of vectors $(q_n,p_n)$,
written in term of the Lax operator $L(x)$.
Candidates for such two-forms are
\beq\label{54}
\omega_{(m)}={1\over 2}\sum_{\a=1}^3{\rm Res}_{P_{\a}}
<\Psi^*_{n+1}(Q) \delta L_n(x)\wedge \delta \Psi_n(Q)>{dx\over x^m}.
\eeq
The various expressions in this equation are defined as follows.
The notation $<f_n>$ stands for a sum over one period of the periodic
function $f_n$:
\beq
<f_n>=\sum_{n=0}^{N+1}f_n.
\eeq
The expression $\psi_n^*(Q)$ is the dual Baker-Akhiezer function, which is
the row-vector solution of the equation
\beq\label{h6}
\psi_{n+1}^*(Q)L_n(z)=\psi_n^*(Q), \ \ \psi_{N+2}^*(Q)=y^{-1}\psi_0^*(Q),
\eeq
normalized by the condition
\beq\label{h7}
\psi_0^*(Q)\psi_0(Q)=1.
\eeq
Note that (\ref{bloch}) and (\ref{h6}) imply that $\psi_{n+1}^*\psi_{n+1}
=\psi_{n+1}^*(L_n(x)\psi_n)
=(\psi_{n+1}^*L_n(x))\psi_n
=\psi_n^*\psi_n$ does not depend on $n$. We would also like
to emphasize that, unlike the Bloch function $\psi_n(Q)$
which does not have $n$-independent zeroes, the normalization (\ref{h7})
allows the dual Bloch function $\psi_n^*(Q)$ to have such zeroes.
In fact, they occur at the poles of $\psi_n(Q)$.

In (\ref{54}), the differential $\delta$ denotes
the exterior differential with respect to the moduli parameters of $\M$.
(This is in order to distinguish $\delta$ from the differential
$d$, which is the exterior differential on the surface $\G$.)
Thus the external differential $\delta L_n(z)$ can be viewed as
a one-form on $\M$, valued in the space of operator-valued
meromorphic functions on $\Gamma$. Similarly the Bloch function
$\psi_n(Q)$ and dual Bloch functions $\psi_n^*(Q)$
are functions on $\cal M$, valued respectively in the space
of column-vector-valued and the space of row-vector-valued
meromorphic functions on $\Gamma$. It follows that
$\delta \psi_n(Q)$ is a one-form on $\M$, valued
in the space of column-vector-valued meromorphic functions
on $\Gamma$. The expression
$\psi_{n+1}^*\delta L_n(x)\wedge \delta\psi_n(x)$ is then a two-form on
$\M$, valued in the space of meromorphic functions on $\Gamma$,
and for each $m$ integer, the expression
\beq
\Omega_{(m)}=<\psi_{n+1}^*\delta L_n(x)\wedge \delta\psi_n(x)>{dx\over
x^m}
\label{0}
\eeq
is a meromorphic 1-form on $\Gamma$. This justifies (\ref{54})
as a two-form on $\M$.

In (\ref{54}), we have allowed for a later choice of an integer $m$.
We shall see shortly that holomorphicity requirements
restrict to $0\leq m\leq 2$, and that the symplectic form of
the $\N=2$ SUSY with a hypermultiplet in the anti-symmetric
representation is obtained by setting $m=0$.

Sometimes it is useful to think of the
symplectic form $\omega$ as
\beq
\omega_{(m)}={1\over 2}{\rm Res}_{x=\infty} \ {\rm Tr}
<\left(\Psi_{n+1}^{-1}(x) \delta L_n(x)\wedge \delta  \Psi_n(x)\right)>
{dx\over x^m},
\label{55}
\eeq
where $\Psi_n(x)$ is a matrix with the columns $\psi_n(Q_j(x)), \
Q_j(x)=(x,y_j)$ corresponding to different sheets of $\G$. The matrix
$\Psi_n(x)$ is of course not defined globally.
Note that $\psi_n^*(Q)$ are the rows of the matrix $\Psi_n^{-1}(x)$.
That implies that $\Psi_n^*(Q)$ as a function on the spectral curve is
meromorphic outside the punctures, has poles
at the branching points of the spectral curve, and zeroes at the poles $z_j$ of
$\Psi_n(Q)$. These analytical properties will be crucial in the sequel.

\subsection{The Symplectic Forms in terms of $x$ and $y$}

A remarkable property of the symplectic form defined by
(\ref{54}) in terms of the Lax operator $L(x)$ is that it
can, under quite general circumstances, be rewritten
in terms of the meromorphic functions $x$ and $y$
on the spectral curve $\Gamma$. More precisely, we have
\beq
\omega_{(m)}=-\sum_{i=1}^{2N+1} \delta \ln y(z_i)\wedge
{\delta x\over x^m}(z_i). \label{61}
\eeq
The meaning of the right hand side of this formula is as follows.
The spectral curve is equipped by definition
with the meromorphic functions $y(Q)$
and $x(Q)$.
Their evaluations $x(z_i),\ y(z_i)$ at the points
$z_i$ define functions on the space $\cal M$, and the wedge product of
their external differentials is a two-form on $\cal M$.

\medskip
The proof of the formula (\ref{61}) is very general
and does not rely on any specific form of $L_n$. For the sake of
completeness we present it here in detail,
although it is very close to the proof of Lemma 5.1
in \cite{k3}.

Recall that the expression $\Omega_{(m)}$ defined in (\ref{0})
is a meromorphic differential on the spectral curve $\G$.
Therefore, the sum of its
residues at the punctures $P_{\a}$
is equal to the opposite of the sum of the other residues on $\Gamma$.
For $m\leq 2$,
the differential $\Omega_{(m)}$ is regular at the points situated over
$x=0$, thanks to the normalization (\ref{nor1}), which insures
that $\delta\psi_n(Q)=O(x)$. Otherwise, it has poles at
the poles $z_i$ of $\psi_n(Q)$ and at the branch points $s_i$,
where we have seen that $\psi_{n+1}^*(Q)$ has poles.
We analyze in turn the residues at each of these two types of poles.

First, we consider the poles $z_i$ of $\psi_n(Q)$. By genericity,
these poles are all distinct and of first order, and we may write
\beq
\psi_n\equiv\psi_{n,0}(z_i){1\over x-x(z_i)}+\cdots
\eeq
It follows that
$\delta \psi_n$ has a pole of second order
at $z_i$
\beq
\delta\psi_n=\psi_{n,0}(z_i){\delta x(z_i)\over (x-x(z_i))^2}+\cdots
\eeq
In view of the fact that $\psi_{n+1}^*$ has a
simple zero at $z_i$ and hence can be expressed as
\beq
\psi_{n+1}^*\equiv\psi_{n+1,0}^*(x-x(z_i))+\cdots,
\eeq
we obtain
\beq
{\rm Res}_{z_i}\Omega_{(m)}=
<\psi_{n+1,0}^*\delta L_n\psi_n>\wedge{\delta x\over x^m}(z_i)
=<\psi_{n+1}^*\delta L_n\psi_n>\wedge {\delta x\over x^m}(z_i).
\label{65}
\eeq
The key observation now is that the right hand side can be
rewritten in terms of the monodromy matrix $L(x)$.
In fact, the recursive relations (\ref{bloch}) and (\ref{h6})
imply that
\begin{eqnarray}\label{651}
<\psi_{n+1}^*\delta L_n\psi_n>
&=&<\psi_{N+2}^*\left(\prod_{m=n+1}^{N+1}L_m\right)
\delta L_n\left(\prod_{m=0}^{n-1}L_m\right)\psi_0>\\
&=&\sum_{n=0}^{N+1}\psi_{N+2}^*\left(\prod_{m=n+1}^{N+1}L_m\right)
\delta L_n\left(\prod_{m=0}^{n-1}L_m\right)\psi_0\\
&=& \psi_{N+2}^*\delta L\psi_0
=\psi_0\delta \ln y\psi_0.
\end{eqnarray}
In the last equality, we have used the standard formula for the
variation of the eigenvalue of an operator,
$\psi_0^*\delta L \psi_0=
\psi_0^*\delta y\psi_0$. Altogether, we have found that
\beq
{\rm Res}_{z_i}\Omega_{(m)}=\delta \ln y(z_i)\wedge {\delta x\over x^m}(z_i).
\label{652}
\eeq

The second set of poles of $\Omega_{(m)}$ is the set of branching points
$s_i$ of the cover. The pole of $\psi_n^*$  at $s_i$ cancels with the zero
of the differential $dx, \ dx(s_i)=0$, considered as a differential on $\G$.
The vector-function $\psi_n$ is holomorphic at $s_i$.
However, $\delta\psi_n$ can develop a pole as we see below.
If we take an expansion of $\psi_n$ in
the local coordinate $(x-x(s_i))^{1/2}$ (in general position when the
branching point is simple) and consider its variation we get
\begin{eqnarray}\label{653}
\psi_n& =& \psi_{n,0}+\psi_{n,\pm}(x-x(s_i))^{1/2}+\cdots\\
\delta\psi_n&= &
-{1\over 2}\psi_{n,\pm}{\delta x(s_i)\over (x-x(s_i))^{1/2}}+\cdots
\end{eqnarray}
Comparing with ${d\psi_n\over dx}={1\over 2}\psi_{n,\pm}{1\over (x-x(s_i))^{1/2}}
+\cdots$, we may write
\beq
\delta \psi_n=-{d\psi_n\over dx}\delta x(s_i)+O(1).\label{66}
\eeq
This shows that $\delta \Psi_n$ has a simple pole at $s_i$. Similarly,
we may write
\beq
\delta y=-{dy\over dx} \delta x(s_i)+O(1). \label{67}
\eeq
The identities (\ref{66}) and (\ref{67}) imply that
\beq
{\rm Res}_{s_i}\Omega_{(m)}=
{\rm Res}_{s_i}\left[ <\psi_{n+1}^*\delta L_n d\psi_n>
\wedge {\delta y \,dx\over x^mdy}\right]\ . \label{68}
\eeq
Arguing as for (\ref{651}), this can be rewritten as
\beq
{\rm Res}_{s_i}\Omega_{(m)}={\rm Res}_{s_i}\left[ \left(\psi_{N+2}^*\delta L
d\psi_0\right)
\wedge {\delta y dx\over x^mdy}\right]\ . \label{681}
\eeq
Due to the antisymmetry of the wedge product, we may replace $\delta L$ in
(\ref{681}) by $(\delta L-\delta y)$. Then using the identities
\begin{eqnarray}
\psi_{N+2}^*(\delta L-\delta y)&= &\delta \psi_{N+2}^* (y-L)\\
(y-L)d\psi_0 &=&(dL-dy)\psi_0,
\end{eqnarray}
which result from $\psi_{N+2}^*(L-y)=(L-y)\psi_0=0$, we obtain
\beq
{\rm Res}_{s_i}\Omega={\rm Res}_{s_i}\left(\delta \psi_{N+2}^*
(dL-dy)\psi_0\right)\wedge
{\delta y dx\over x^m dy}
\eeq
Now the differential $dL$ does not contribute to the residue, since
$dL(s_i)=0$. Furthermore,
$\psi_{N+2}^*\psi_0=y^{-1}\psi_0^*\psi_0=y^{-1}$.
Thus $\delta\psi_{N+2}^*\psi_0=-\psi_{N+2}^*\delta\psi_0-y^{-2}\delta y$.
Exploiting again the antisymmetry of the wedge product,
we arrive at
\beq
{\rm Res}_{s_i}\Omega=
{\rm Res}_{s_i}\left(\psi_{N+2}^*\delta \psi_0\right)\wedge
 \delta y {dx\over x^m}. \label{000}
\eeq

Recall that we have normalized the Bloch function $\psi_0(Q)$
at $x=0$ by (\ref{nor1}), and that near $x=0$, the function
$y$ is of the form (\ref{100}). Thus $\delta\psi_0= O(x)$
and $\delta y=O(x)$ near $x=0$, and the differential form
\beq
\left(\psi_{N+2}^*\delta \psi_0\right)\wedge \delta y {dx\over x^m}.
\label{001}
\eeq
is holomorphic at $x=0$ for $0\leq m\leq 2$.
It is manifestly holomorphic at all the other points of $\G$,
except at the branching points $s_i$ and
the poles $z_1,\cdots,z_{2N+1}$.
Therefore
\beq\label{527}
\sum_{s_i}{\rm Res}_{s_i}\left(\psi_{N+2}^*\delta \psi_0\right)
\wedge \delta y {dx\over x^m}=
-\sum_{i=1}^{2N+1}{\rm Res}_{z_i}\left(\psi_{N+2}^*\delta \psi_0\right)
\wedge \delta y {dx\over x^m}
\eeq
Using again the expressions (\ref{653}, \ref{66})
for $\psi_0$ and $\delta\psi_0$,
and the fact that $\psi_{N+2}^*=y^{-1}\psi_0^*$,
the right hand side of (\ref{527}) can be recognized as
\beq\label{h10}
\sum_{i=1}^{2N+1}\delta \ln y(z_i)\wedge {\delta x(z_i)\over x^{m}(z_i)}.
\eeq
The sum of (\ref{652}) and (\ref{h10}) gives (\ref{61}), since
\beq
2\omega_{(m)}=-\sum_{i=1}^{2N}{\rm Res}_{z_i}\Omega_{(m)}-\sum_{s_i}{\rm
Res}_{s_i}\Omega_{(m)}.
\eeq
The identity (\ref{61}) is proved.

\subsection{Action-Angle Variables
and Seiberg-Witten Differential}

The expression (\ref{61}) for the symplectic form $\omega_{(m)}$
suggests its close relation with the following one-form on $\G$
\beq
d\lambda_{(m)}=\ln y\,{dx\over x^m}\label{dl}
\eeq
Strictly speaking, the form $d\lambda_{(m)}$ is not
a meromorphic differential in the usual sense, because of the
multiple-valuedness of $\ln \,y$. However, the ambiguities in $\ln\,y$
are fixed multiples of $2\pi i$, which disappear upon
differentiation. Thus, the form $d\lambda_{(m)}$ is no different
from the usual meromorphic differentials, as far
as the construction of symplectic forms is concerned. Also,
the form $d\lambda_{(m)}$ and the form ${1\over m-1}x^{-m+1}{dy\over y}$
(for $m\not=1$; for $m=1$, $-(\ln \,x){dy\over y}$)
differ by an exact differential, and we shall not distinguish
between them. From this point of view, the Seiberg-Witten form
(\ref{sw1}) can be identified with the form $-d\lambda_{(0)}$.

Our spin chain model has led so far to a $2N$-dimensional
phase space $\M$, equipped with several candidate symplectic
forms $\omega_{(m)}$, $1\leq m\leq 2$. We still have to reduce
$\M$ to a $(2N-2)$-dimensional phase space, and to identify the
correct symplectic form. Remarkably, both selections are tied to
a key physical requirement for the one-form which corresponds
to the Seiberg-Witten of a $\N=2$ SUSY gauge theory,
namely the holomorphicity of its variations under moduli
deformations.

It is an important feature of $\N=2$ Yang-Mills theories
that the masses of the theory are not renormalized.
Since the masses of the theory correspond to the poles of
the Seiberg-Witten differential $d\lambda$, it follows that
$\delta d\lambda$ must be holomorphic.
Thus we need to examine
the poles of $\delta\,d\lambda=\delta\ln y {dx\over x^m}$,
and identify the subvarieties of $\M$ along
which $\delta d\lambda$
is holomorphic. There are 3 such subvarieties, corresponding
to the choices of $m$:

\medskip

$\bullet$
On the variety ${\M}_2=\M\cap\{u_0=c_0,\ u_1=c_1\}$, the differential
$\delta\, d\lambda_{(2)}=(\delta \ln y){dx\over x^2}$ has no pole at $Q_{\a}$,
since $y=1+O(x^2)$ near $x=0$.
On the other hand, the differential ${dx\over x^2}$ vanishes at $x=\infty$,
so $\delta\ln y{dx\over x^2}$ is also holomorphic there,
and $\delta d\lambda_{(2)}$ is holomorphic.

\medskip

$\bullet$
On the variety ${\M}_0=\M\cap\{u_N=1,\ u_{N-1}=0\}$, the differential
$\delta\, d\lambda_{(0)}=(\delta\ln y)dx$ is automatically holomorphic
at $x=0$. Near $\infty$, in view of the expansion ()
for $y$, we have $\delta\ln y=O(x^2)$ if we vary
only the moduli within $\M_2$. Thus $\delta d\lambda_{(0)}$
is holomorphic.

\medskip

$\bullet$
On the variety ${\M}_1=\M\cap\{u_N=1\}$, the differential $\delta\,d\lambda_{(1)}
=(\delta\ln y){dx\over x}$ is still holomorphic, because $\delta\,\ln y
=O(x)$. Near $x=\infty$, the sole constraint $\{u_{N-1}=1\}$
suffices to guarantee that $\delta\ln y=O({1\over x})$.
Thus $\delta d\lambda_{(1)}$ is holomorphic.

\medskip

When $m$ and hence $d\lambda_{(m)}$
is even under the involution $\sigma$,
action-angle variables can be introduced as follows.
Restricted to $\M_{(m)}$, $\delta d\lambda_{(m)}$ is holomorphic,
and hence can be expressed for suitable coefficients
$\delta a_i$ as
\beq\label{deltaa}
\delta d\lambda_{(m)}=\sum_{i=1}^{2N-1}(\delta a_i)d\omega_i,
\eeq
where $d\omega_i$ is a basis of $2N-1$ holomorphic one-forms on $\G$.
Since $d\lambda_{(m)}$ is even, only holomorphic one-forms $d\omega_i$
which are even can occur on the right hand side. We identify such
forms with forms on $\G/\sigma$.
We choose a symplectic homology basis $A_i,B_i$
and a dual basis of holomorphic forms $d\omega_i$,
$1\leq i\leq N-1$, for the factor curve $\Gamma/\sigma$.
The variables $a_i$ and $a_{Di}$ can then be defined by
\beq\label{a}
a_i=\oint_{A_i}d\l_{(m)},
\quad
a_{Di}=\oint_{B_i}d\l_{(m)}.
\eeq
The interpretation of the variables $a_i$ is as action variables
from the viewpoint of the spin model and as vacuum moduli from the
viewpoint of the $\N=2$ SUSY
gauge theory. Evidently, their variations
coincide with the $\delta a_i$ of the equation (\ref{deltaa}).

Next, the angle variables $\phi_i$, $1\leq i\leq N-1$,
are defined by
\beq\label{a5}
D=\{z_1,\cdots,z_{2N+1}\}
\longmapsto \phi_i=\sum_{j=1}^{2N+1} \int^{z_j}d\omega_i
\eeq
We claim now that, for $m$ even, the symplectic form $\omega_{(m)}$
is a genuine symplectic form when restricted to $\M_{(m)}$,
and that $a_i$ and $\phi_i$ as defined above are action-angle
coordinates for $\omega_{(m)}$
\beq\label{a6}
\omega_{(m)}=\sum_{i=1}^{N-1}\delta a_i\wedge \delta \phi_i
\quad
{\rm on}
\ \M_{(m)}.
\eeq
To see this, we evaluate the two-form
$\delta\big(\sum_{j=1}^{2N+1}\int_{Q_0}^{z_j}\delta\,d\lambda\big)$
in two different ways. Substituting in (\ref{deltaa}),
we find that it is equal to
\beq
\delta(\sum_{i=1}^{N-1}\delta a_i\,\phi_i)
=\sum_{i=1}^{N-1}\delta \phi_i\wedge\delta a_i.
\eeq
On the other hand, we can also write
\beq
\delta\left(\sum_{j=1}^{2N+1}\int_{Q_0}^{z_j}\delta\,d\lambda\right)=
\delta\left(\sum_{j=1}^{2N+1}\int_{Q_0}^{z_j}
(\delta\ln\,y){dx\over x^m}\right)=
\sum_{j=1}^{2N+1}{\delta x(z_j)\over x^m(z_j)}\wedge (\delta\ln \,y)(z_j).
\eeq
Comparing the two formulas, and making use of (\ref{61}),
we obtain the desired equation (\ref{a6}).

We observe that for the present even divisor spin model,
the space $\M_1$ and the form $d\lambda_{(1)}$ are not applicable.
In fact, there are difficulties with both
the dimension of $\M_1$ which is odd, and the angle variables
$\phi_i$ defined by (\ref{a5}), which would vanish
identically because the class of the divisor $D$ is even.

For the $\N=2$ SUSY Yang-Mills theory with a hypermultiplet
in the antisymmetric representation, the spectral curves are given
by $\M_0$. The symplectic form is then $\omega_{(0)}$,
which provides an independent check of the choice of
Seiberg-Witten form found by Landsteiner and Lopez.

\subsection{The Hamiltonian of the Flow}

We show now that the even divisor spin model
is a Hamiltonian system. More precisely, restricted to each
of the phase space $\M_{(0)}$ or $\M_{(2)}$,
the system is Hamiltonian with the corresponding
symplectic form, with a corresponding Hamiltonian.
We would like to stress that, once again, the arguments
to these ends are quite general, and use only the
expression for $\omega_{(m)}$ in terms of the Lax operator.

\begin{lem} Let $m$ be either $0$ or $2$.
Then the equations (\ref{pq0}) restricted on $\M_{(m)}$ are
Hamiltonian with respect to the symplectic form $\omega_{(m)}$ given
by (\ref{54}). The
Hamiltonians $H_{(m)}$ are given by
\begin{eqnarray}\label{a7}
H_{(0)}&=& u_{N-2}\\
H_{(2)}&=&\ln u_N=\sum_{n=0}^{N+1}\ln (p_n^+q_{n-1})=
{1\over 2}\sum_{n=0}^{N+1}\ln [(p_n^+q_{n-1})(p_{n-1}^+q_{n})]
\end{eqnarray}
\end{lem}

\noindent{\it Proof.}
By definition, a vector field $\p_t$ on a symplectic
manifold is Hamiltonian, if its contraction $i_{\p_t}\omega(X)=
\omega(X,\p_t)$ with the symplectic form
is an exact one-form $\delta H(X)$. The function $H$ is the Hamiltonian
corresponding to the vector field $\p_t$. Thus
\beq
i_{\p_t}\omega_{(m)}={1\over 2}\sum_{\a}{\rm Res}_{P_{\a}}
\left(<\psi_{n+1}^*\delta L_n\dot\psi_n>-<\psi_{n+1}^*
\dot L_n\delta \psi_n> \right){dx\over x^m}
\eeq
Now under the flow (\ref{pq0}), the Lax operators $L_n(x)$
flow according to the Lax equation
(\ref{laxLM}), while the Bloch function $\psi_n$ flow
according to (\ref{12}). Consequently,
\beq\label{ham}
i_{\p_t}\omega_{(m)}={1\over 2}\sum_{\a}{\rm Res}_{P_{\a}}
\left(<\psi_{n+1}^*\delta L_n(M_n+\mu)\psi_n>-<\psi_{n+1}^*
(M_{n+1}L_n-L_nM_n)\delta \psi_n> \right){dx\over x^m}.
\eeq
Since $L_n\psi_n=\psi_{n+1}$,
it follows that $\psi_{n+1}^*M_{n+1}L_n\delta\psi_n
=\psi_{n+1}^*M_{n+1}\psi_{n+1}-\psi_{n+1}^*M_{n+1}\delta L_n\psi_n$.
Upon averaging in $n$, we obtain
\beq
<\psi_{n+1}^*(M_{n+1}L_n-L_nM_n)\delta\psi_n>
=
-<\psi_{n+1}^*M_{n+1}\delta L_n\psi_n>
\eeq
For all $n$, both $\delta L_n(x)$
and $M_n(x)$ vanish at $x=0$. The differential form
\beq
<\psi_{n+1}^*
\left(\delta L_nM_n+M_{n+1}\delta L_n\right)\psi_n>{dx\over x^m}
\eeq
is thus holomorphic at $x=0$, in both cases $m=0$ and $m=2$.
As we have seen, outside of $x=\infty$,
the poles of $\psi_{n+1}^*$ are at the branch
ponits and are cancelled by the zeroes of $dx$ there,
while the poles of $\psi_n$ are cancelled by the zeroes
of $\psi_{n+1}^*$. Thus the above differential form
is holomorphic outside of $x=0$.
The sum of its residues at $P_{\a}$ must be zero
\beq
\sum_{\a}{\rm Res}_{P_{\a}}<\psi_{n+1}^*
\left(\delta L_nM_n+M_{n+1}\delta L_n\right)\psi_n>{dx\over x^m}=0.
\eeq
The expression (\ref{ham}) for $i_{\p_t}\omega_{(m)}$
reduces to
\beq
i_{\p_t}\omega_{(m)}={1\over 2}
\sum_{\a}{\rm Res}_{P_{\a}} \left(<\psi_{n+1}^* \delta L_n\psi_n>
\mu(Q,t)\right){dx\over x^m}
\eeq
Applying the arguments leading to (\ref{651}), we obtain
\beq\label{ham1}
i_{\p_t}\omega_{(m)}
=
{1\over 2}\sum_{\a}\res_{P_{\a}}  \delta(\ln y)
\mu(t,Q){dx\over x^m}\ .
\eeq
As follows from (\ref{ex1},\ref{ex11}), and (\ref{12}) the function $\mu(t,Q)$
is holomorphic at $P_2$, while it has the following
expansion at the punctures $P_1, \ P_3$
\beq
\mu(t,Q)=-x+O(1), \ Q\to P_1 ;\ \ \mu(t,Q)=x+O(1), \ Q\to P_3.
\eeq
We consider now the cases $m=2$ and $m=0$ separately.
When $m=2$, the form $\mu{dx\over x^2}$ is regular at $P_2$,
and has simple poles with opposite residues at $P_1$ and $P_3$.
Since $\delta \ln y=\delta u_N+O({1\over x})$ near $P_1$,
it follows immediately that
\beq
i_{\p_t}\omega_{(2)}=\delta(\ln u_N).
\eeq
When $m=0$, we observe that
the form $(\delta\,\ln \,y) dx$ is regular at $x=\infty$.
Indeed, the constraints $u_N=1$, $u_{N-1}=0$ defining the phase
space $\M_0$ in this case imply that $\delta\ln y=O({1\over x^2})$
near all three points $P_1,P_2$, and $P_3$.
For $P_1$ and $P_3$, this statement is a direct consequence of
(\ref{y27}) and (\ref{y28}). For $P_2$, this follows from the fact
that the three roots $y_{\a}$ of the Landsteiner-Lopez curve
(\ref{ll1}) must satisfy $\prod_{\a=1}^3y_{\a}=1$.
Returning to the residues in (\ref{ham1}), we see that the point
$P_2$ does not contribute. As for the points $P_1$ and $P_3$,
they contribute exactly the coefficient $u_{N-2}$
in the expansions (\ref{y27}) and (\ref{y28}) for $y$
\beq
i_{\p_t}\omega_{(0)}=\delta u_{N-2}.
\eeq
The lemma is proved.

\subsection{The Symplectic Form in Terms of $(p_n,q_n)$}

The expression (\ref{54}) for the symplectic forms $\omega_{(m)}$
in terms of the Lax operator also provides a straightforward
way of writing $\omega_{(m)}$ in terms of the dynamical
variables $(q_n,p_n)$. Such an expression for the form $\omega_{(0)}$
appears complicated. But it is quite simple for the form $\omega_{(2)}$,
and we derive it here.

We have $\delta L_n=x\,\delta (q_np_n^T)$, and the contributions
of the three points $P_a$ above $x=\infty$ can be evaluated
as follows.

\medskip

At the point $P_1$, $y=O(x^{N+2})$, $\psi_n=O(x^n)$,
$\psi_{n+1}=O(x^{-(n+1)})$, and thus the differential
$<\psi_{n+1}^*\delta L_n\wedge\delta\psi_n>{dx\over x^2}$
is regular. The residue at $P_1$ vanishes.

\medskip
At the point $P_2$, $\psi_n$ and $\psi_{n+1}^*$ are regular.
Using the same notation as in (\ref{expansion}),
we write
\begin{eqnarray}
\psi_n&=&\psi_{n,0}+\psi_{n,1}x^{-1}+\cdots\\
\psi_{n+1}^*
&=&\psi_{n+1,0}^*+\psi_{n+1,1}^*x^{-1}+\cdots
\end{eqnarray}
In analogy with (ref),
from the equation
\beq
\psi_{n+1}^*=\psi_n^*L_n(x)^{-1}
=\psi_n^*(1-x q_np_n^T),\label{psistar}
\eeq
it follows that
\beq
\psi_{n,0}^*q_n=\psi_{n+1,0}^*q_n=0
\eeq
The residue at $P_2$ is then readily identified
\begin{eqnarray}
{\rm Res}_{P_2}<\psi_{n+1}^*\delta L_n\wedge\delta\psi_n>{dx\over x^2}
&=&
{\rm Res}_{P_2}<\psi_{n+1,0}^*\delta(q_np_n^T)\wedge \delta\psi_{n,0}>
{dx\over x}\\
&=&
-<\psi_{n+1,0}^*\delta q_n\wedge (\delta p_n^T)\psi_{n,1}>\\
&\equiv & I.
\end{eqnarray}

At the point $P_3$, $y=O(x^{-N-2})$, and
\begin{eqnarray}
\psi_n &=& \psi_{n,0}x^{-n}+\psi_{n,1}x^{-n-1}+\cdots\\
\psi_{n+1}^*&=& \psi_{n+1,0}^*x^{n+1}+
\psi_{n+1,1}^*x^n+\cdots
\end{eqnarray}
It follows that the residue is given by
\beq
{\rm Res}_{P_3}<\psi_{n+1}^*\delta L_n\wedge\delta\psi_n>{dx\over x^2}
=
\big[\psi_{n+1,0}^*\delta(q_np_n^T)\wedge \delta\psi_{n,1}
+
\psi_{n+1,1}^*\delta(q_np_n^T)\wedge \delta\psi_{n,0}\big]
\eeq
We now make use of the equation (\ref{psistar}) to derive
recursion relations between the coefficients of $\psi_n^*$
\beq
\psi_{n+1,0}^*=-\psi_{n,0}^*q_np_n^T,
\quad
\psi_{n+1,1}^*
=
\psi_{n,0}^*-\psi_{n,1}^*q_np_n^T.\label{recursion2}
\eeq
They imply that
\beq
\psi_{n+1,0}^*q_n=0,
\quad
\psi_{n+1,1}^*q_n=\psi_{n,0}^*q_n
\eeq
As a consequence, the first term on the right hand side of ()
simplifies to
\beq
\psi_{n+1,0}^*\delta(q_np_n^T)\wedge\delta\psi_{n,1}
=\psi_{n+1,0}^*\delta q_n\wedge p_n^T\delta\psi_{n,1}
\eeq
Now recall that we introduced the coefficient $\beta_n$
by $\psi_n=\beta_n q_n$. Comparing with the equation (),
we obtain
\beq
\beta_n=-p_n^T\psi_{n,1}
\eeq
and the preceding term becomes
\beq
\psi_{n+1,0}^*\delta(q_np_n^T)\wedge\delta\psi_{n,1}
=
-\psi_{n+1,0}^*\delta q_n\wedge \delta\beta_n
-
\psi_{n+1,0}^*(\delta q_n\wedge\delta p_n^T)\psi_{n,1}
\eeq
On the other hand, $p_n^T\psi_{n,0}=0$, and the second term
on the right hand side of () can be rewritten as
\beq
\psi_{n+1,1}^*\delta(q_np_n^T)\wedge \delta\psi_{n,0}
=
\psi_{n+1,1}^*q_n\delta p_n^T\wedge \delta\psi_{n,0}
-
\psi_{n+1,1}^*\delta q_n\wedge (\delta p_n^T)\psi_{n,0}
\eeq
Altogether, we obtain the following expression for the residue
at $P_3$
\beq
{\rm Res}_{P_3}<\psi_{n+1}^*\delta L_n\wedge\delta\psi_n>{dx\over x^2}
=
{\rm II+III}
\eeq
where the terms II and III are defined by
\begin{eqnarray}
{\rm II}
&=&
-[\psi_{n+1,0}^*(\delta q_n\wedge \delta p_n^T)\psi_{n,1}
+
\psi_{n+1,1}^*(\delta q_n\wedge \delta p_n^T)\psi_{n,0}]\\
{\rm III}&= &
-(\psi_{n+1,0}^*\delta q_n\wedge\delta\beta_n
-
\psi_{n+1,1}^*q_n\delta p_n^T\wedge \delta \psi_{n,0})
\end{eqnarray}
We claim that the term III can be simplified to
\beq
{\rm III}=-\delta p_n^T\wedge\delta q_n\label{III}
\eeq
In fact, in view of the recursion relations (\ref{recursion2})
and the fact that $\psi_n=\beta_nq_n$, it can
be rewritten as
\beq
{\rm III}=-\psi_{n,0}^*q_n)\big(
p_n^T\delta q_n\wedge \delta\beta_n+
\delta p_n^T\wedge(\delta \beta_n)q_n
+
\delta p_n^T\wedge\beta_n\delta q_n\big)
\eeq
The first two terms on the right hand side cancel,
since $p_n^Tq_n=0$. As for the remaining term, we note that
the normalization $\psi_n^*\psi_n=1$ implies near $P_3$
\beq
1=(\psi_{n,0}^*x^{-n}+O(x^{-n-1}))(\beta_nq_nx^n+O(x^{n-1}))
=\psi_{n,0}^*\beta_nq_n+O(x^{-1})
\eeq
from which it follows that $\psi_{n,0}^*\beta_nq_n=1$.
The identity (\ref{III}) is established.

Finally, it is readily seen that the remaining terms I and II
combine into
\beq
{\rm I}+{\rm II}=-\sum_{a=1}^3{\rm Res}_{P_a}<\psi_{n+1}^*\delta q_n\wedge
\delta p_n^T\psi_n>{dx\over x}
\eeq
But the 1-form $<\psi_{n+1}^*\delta q_n\wedge
\delta p_n^T\psi_n>{dx\over x}$ is meromorphic on the space $\Gamma$,
with poles only at the points $P_a$ above $x=\infty$ and
$Q_a$ above $x=0$. We can deform then contours and rewrite II+III
as residues at $Q_a$
\beq
{\rm I+II}
=\sum_{a=1}^3{\rm Res}_{Q_a}<\psi_{n+1}^*\delta q_n\wedge
\delta p_n^T\psi_n>{dx\over x}
\eeq
At $x=0$, we have $\psi_{n+1}^*=\psi_n^*$,
and this expression is determined by the normalization condition
(\ref{nor1}) on the matrix $W$. In terms of $\psi_n$,
the normalization (\ref{nor1}) can be restated as the normalization
condition $\psi_n^*(0)\psi_n^T=I$ as an identity
between $3\times 3$ matrices. Thus ${\rm I+II}
=3<\delta q_n\wedge\delta p_n>$, and we obtain the final
formula for the symplectic form $\omega$ in terms of $p_n$ and $q_n$
\beq
\omega
=
2\sum_{n=0}^{N+1}\delta q_n^T\wedge\delta p_n.
\eeq

\section{Hamiltonian theory and Seiberg-Witten differential:
The Odd Divisor Model}

The main difference between the even and the odd divisor spin models
is in the parity of the divisor $D$ of poles of the Bloch function
$\psi_n(Q)$. For the odd divisor spin model, $D$ is essentially odd
under the involution $\sigma:(x,y)\to (-x,y^{-1})$
in the following sense
\beq
[D]+[D^{\sigma}]
=K+2\sum_{\alpha=1}^3P_{\alpha}\label{oddd}
\eeq
Here $K$ is the canonical class, which is the divisor class of
any meromorphic 1-forms on $\Gamma$. As in the case of the even divisor
spin model, the relation (\ref{oddd}) is a consequence of the
transformation of $L(x)$ under $\sigma$, which is in this
case $L(-x)=(L(x)^{-1})^T$. This implies that $\psi_0(Q^{\sigma})$
and $\psi_0(Q)^*$ are both dual Bloch functions for $L(x)$,
and thus
\beq
\psi_0^*(Q)=\psi_0(Q^{\sigma})f(Q)
\eeq
where $f(Q)$ is a meromorphic function on $\Gamma$.
But the zeroes of the dual Bloch function $\psi_0^*$ are exactly
the poles of $\psi_0(Q)$, while its poles are exactly
the branch points of the surface $\Gamma$. Thus the preceding
equation implies the following equation for divisor classes
\beq
[{\rm branch\ points}]
-[D]=[D^{\sigma}]
\eeq
To determine the divisor of the branch points of $\Gamma$,
we consider the differential $dx$, viewed as a meromorphic
form on $\Gamma$. Since $dx$ has a pole of order $2$ at each $P_a$,
and a zero at each branch point, we have $[{\rm branch\ points}]
-2\sum_{a=1}^3P_a=K$, and the desired relation (\ref{oddd})
follows.

$\bullet$ We discuss briefly the direct and the inverse problems for
the odd divisor spin system. Once the difference in parity of the
divisor of poles of the Bloch functions is taken into account,
the direct problem is treated in exactly the same way as before.
As for the inverse problem, we need only a few minor modifications
in expansions near the punctures $P_1,P_3$, which we give (c.f. (\ref{ex1},
\ref{ex11}) now
\beq
\phi_n=x^ne^{xt}\left(\sum_{k=0}^{\infty}\phi_{n,k}(P_1)x^{-k}\right),\ \ Q\to P_1,
\label{ex16}
\eeq
\beq
\phi_n=x^{-n}e^{xt}\left(\sum_{k=0}^{\infty}\phi_{n,k}(P_3)x^{-k}\right),\ \ Q\to P_3,
\label{ex116}
\eeq
They lead to minor modifications in the exact formulas for the
Baker-Akhiezer function $\phi_n(t,Q)$ (c.f. (\ref{BA})):
\beq\label{BA1}
\phi_{n,\a}(t,Q)=r_{\alpha}(Q)
\frac{\theta(A(Q)+tU^-+nV+Z_{\alpha})\;\theta(Z_0)}{%
\theta(A(Q)+Z_{\alpha})\;\theta(tU^-+nV+Z_0)}
\exp{\left(\int_{Q_{\alpha}}^Q nd\Omega_0+td\Omega^-\right)}
\eeq
where $d\Omega^-=d\Omega_1-d\Omega_1^{\s}$ and $U^-=U-U^{\s}$.

We show that if the divisor $D$ satisfies (\ref{oddd}), then the
corresponding Baker-Akhiezer function satisfies the relation
\beq\label{611}
\phi_n^*(t,Q)=\phi_n^T(t,Q^{\s})f(Q),
\eeq
where as before $\phi_n^*$ are the rows of the matrix inverse to
the matrix
\beq\label{622}
\Phi_{n,\a}^{\b}(x)=\phi_{n,\a}(P_{\b})
\eeq
Here the points $P_{\a}(x)$ are the three preimages of $x$ on $\G$
on different sheets.
Of course, the matrix $\Phi_n(x)$ does depend on the ordereing of sheets,
but one can check that if for $P_{\a}(x)$ we define $\phi_n^*(P_{\a})$ as
the corresponding row of the inverse matrix, then $\phi_n^*$ is well-defined.
As before $\phi_n^*$ has poles at all the branching points and zeroes
at the points of the divisor $D$.

To establish (\ref{611}), we show that
\beq
\sum_{\a}\phi_{n,\a}(t,P_{\g}(x))\phi_{n,\b}^{\s}(t,P_{\g})f(P_{\g})=
\delta_{\a,\b}
\eeq
Indeed, from (\ref{ex16}) and (\ref{ex116}), it follows that the function
$\phi_{n,\a}(t,Q)\phi_{n,\b}(Q^{\s})f(Q)$ is holomorphic everywhere
except at the branching points (the poles and the essential singularities at the
punctures $P_{\a}$ over $x=\infty$ cancel each other; there are no poles
at $D$ and $D^{\s}$ because $f(Q)$ has zeros at these points).
Therefore, the left hand side of the above equation is a
holomorphic function of $x$ (the poles at
the branching points cancel upon the summation). Hence it is a constant,
which can be found by taking $x=0$.

The uniqueness of $\phi_n$ and the relation (\ref{611})
implies as before that it satisfy
the equation
\beq\label{63}
\phi_{n+1}=L_n(x)\phi_n , \p_t \phi_n=M_n(x)\phi_n
\eeq
where $L_n$ and $M_n$ have the form (\ref{laxoddl},\ref{laxodd}).

$\bullet$ We come now to the Hamiltonian structure of the odd divisor
spin model. Recall that we had introduced the space $\M^{\rm odd}$ of
spin chains. Solving the direct and inverse spectral problem
as in the case of the even divisor spin model,
we can identity $\M^{\rm odd}$ with the space of geometric
data
\beq
{\cal M}_1^{\rm odd}
\leftrightarrow
\{\Gamma,D;\ [D]+[D^{\sigma}]=K+2\sum_{\a=1}^3P_{\a}\}
\eeq
We can verify that the space on the right hand side is $2N+1$
dimensional, as it should be: there are
$N+1$ moduli parameters for the curve $\Gamma$,
and $N$ parameters for the antisymmetric divisor $[D]$.
The same discussion as in \S 5.3 and \S 5.4
for the even divisor
spin model shows that, in the present case, the only candidate
for symplectic form is the form $\omega_{(1)}$, restricted
to the $2N$-dimensional phase space $\M_1^{\rm odd}$ defined by
\beq
\M_1^{\rm odd}=\M^{\rm odd}\cap\{u_N=1\}
\eeq
The corresponding action and angle variables are now given by
\beq
a_i=\oint_{A_i^{\rm odd}}d\lambda_{(1)},
\quad
\phi_i=\sum_{j=1}^{2N+1}\int^{z_j}d\omega_i^{\rm odd},
\quad
1\leq i\leq N
\eeq
where $d\omega_i^{\rm odd}$
and $A_i^{\rm odd}$ are respectively a basis of odd holomorphic
differentials and a basis of odd $A$-cycles. We have then as before
\beq
\omega_{(1)}=\sum_{j=1}^N\delta a_j\wedge\delta \phi_j.
\eeq
\vfill\break

\end{document}